\documentclass{aa}  

\usepackage{graphicx}
\usepackage{txfonts}
\usepackage{lipsum}
\usepackage{subcaption}         
\usepackage{lscape}             
\usepackage{placeins}           
                                
\begin{document}

   \title{ALMA 873 $\mu$m Polarization Observations of the PDS~70 Disk}


%

   \author{Hauyu Baobab Liu\inst{1,2}\fnmsep\thanks{Corresponding author: hyliu.nsysu@g-mail.nsysu.edu.tw}
        \and Kiyoaki Doi\inst{3,4}
        \and Simon Casassus\inst{5,6,7}
        \and Akimasa Kataoka\inst{4}
        \and Ruobing Dong\inst{8}
        \and Jun Hashimoto\inst{9}
        \and Philipp Weber\inst{10,11,12}
        }

   \institute{Department of Physics, National Sun Yat-Sen University, No. 70, Lien-Hai Road, Kaohsiung City 80424, Taiwan, R.O.C.
   \and Center of Astronomy and Gravitation, National Taiwan Normal University, Taipei 116, Taiwan
   \and Department of Astronomical Science, School of Physical Sciences, Graduate University for Advanced Studies (SOKENDAI), 2-21-1 Osawa, Mitaka, Tokyo 181-8588, Japan
   \and National Astronomical Observatory of Japan, 2-21-1 Osawa, Mitaka, Tokyo 181-8588, Japan
   \and Departamento de Astronom\'ia, Universidad de Chile, Casilla 36-D, Santiago, Chile
   \and Data Observatory Foundation, to Data Observatory Foundation, Eliodoro Ya\'n\~ez 2990, Providencia, Santiago, Chile
   \and Millennium Nucleus on Young Exoplanets and Their Moons (YEMS), Santiago, Chile
   \and Kavli Institute for Astronomy and Astrophysics, Peking University, 5 Yiheyuan Road, Haidian District, Beijing, 100871, People's Republic of China
   \and Institute of Astronomy and Astrophysics, Academia Sinica, 11F of Astronomy-Mathematics Building, AS/NTU No.1, Sec. 4, Roosevelt Rd, Taipei 10617, Taiwan, R.O.C.
   \and Departamento de Física, Universidad de Santiago de Chile, Av. V\'ictor Jara 3493, Santiago, Chile
   \and Millennium Nucleus on Young Exoplanets and their Moons (YEMS), Chile
   \and Center for Interdisciplinary Research in Astrophysics Space Exploration (CIRAS), Universidad de Santiago, Chile
   }

   \date{Received December 23, 2025}

 

\abstract{
At a 112.4 pc distance, the PDS\,70 protoplanetary disk is a rare case that has been confirmed to host two accreting planets.
This makes it the most important laboratory for studying dust growth in the context of planet formation.
Here we present the first deep, full polarization observations at 873\,$\mu$m wavelength.
We detected $\sim$1\%--2.5\% linear polarization over the bulk of the $\sim$55--100 AU (sub)millimeter ring.
The polarization position angles align preferentially with the projected minor axis of the disk.
The standard interpretation is that the observed polarization is caused by dust self-scattering, with a maximum dust grain size of $\sim$100 $\mu$m. 
On $\gtrsim$10 AU scales, which can be resolved by the presented 873--3075 $\mu$m observations, the ring is marginally optical thick at 873 $\mu$m wavelength.
Using Monte Carlo radiative transfer simulations, we found that an azimuthally asymmetric, marginally optically thick ring with a maximum dust grain size of $\sim$87 $\mu$m can reproduce the observed 873 $\mu$m polarization position angles and percentages. 
This study indicates that the coagulation of ice-coated dust in the protoplanetary disk may be limited by fragmentation or bouncing.
}


   \maketitle
\nolinenumbers

\section{Introduction}

For a region in the protoplanetary disk, formation of rocky planet is promoted if dust grains can grow to $\gtrsim$10$^{3}$\,$\mu$m sizes by coagulation.
Once the dust grains reach this size, they can become partially dynamically decoupled from the surrounding gas and begin to drift toward pressure maxima \citep{Weidenschilling1977MNRAS.180...57W}, which results in local dust overdensities.
They can manifest as substructures in the protoplanetary disk.
Inside these dust overdensities, various dynamical instabilities tend to make evolved dust spatially much more confined than gas.
In the end, some of these dust concentrations may undergo self-gravitational contraction, resulting in the formation of kilometer-sized planetesimals \citep{Birnstiel2016SSRv..205...41B}. 

Why may grains grow to $\gtrsim$10$^{3}$\,$\mu$m in certain regions in a protoplanetary disk, but not necessarily everywhere? 
This is, to a large extent, because the stickiness of dust surface depends on the chemical composition, which depends on temperature. 
Inside the $\sim$150 K isotherm, known as the water snowline, dust grains are primarily composed of astronomical silicate and carbonaceous material.
In contrast, dust grains located outside the water snowline are coated with significant amount of water ice.
Most of the area in a protoplanetary disk is situated outside of the water snowline. 
Whether or not water-ice-coated dust grains are sticky enough to coagulate to $\gtrsim$10$^{3}$ $\mu$m sizes efficiently, remain  actively debated \citep{Musiolik2019ApJ...873...58M,Kimura2020MNRAS.498.1801K}.
Moreover, planets that formed in low-temperature environments likely have high abundances of water and carbon. 
This poses challenges to reconciling with observations of our own Earth, where the carbon and water abundances are only about $\sim$0.1\% of those found in interstellar dust \citep{Li2021SciA....7.3632L}.

\begin{figure*}
    \begin{center}
    \begin{tabular}{ c c }
    \includegraphics[height=8.5cm]{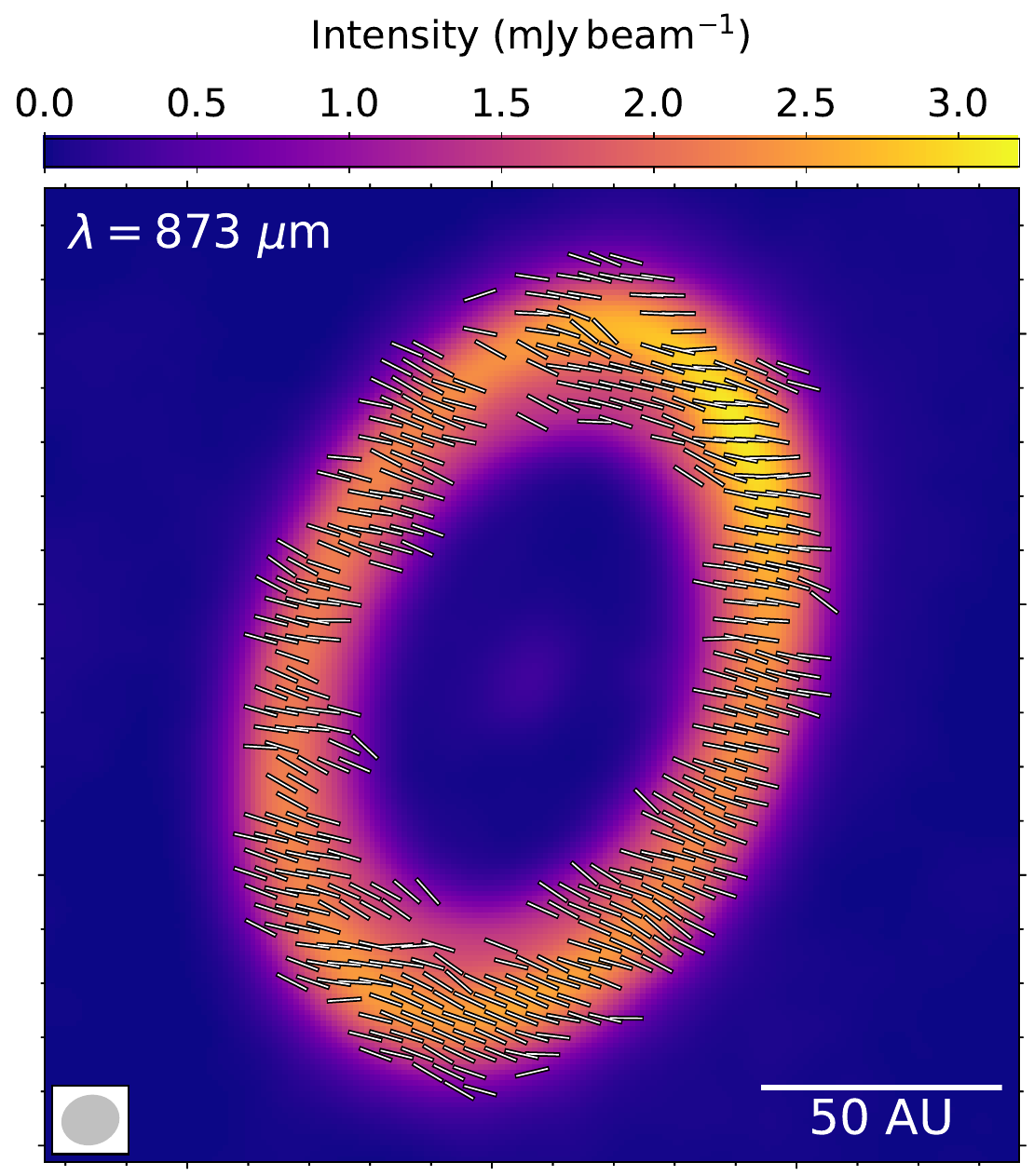} &
    \includegraphics[height=8.5cm]{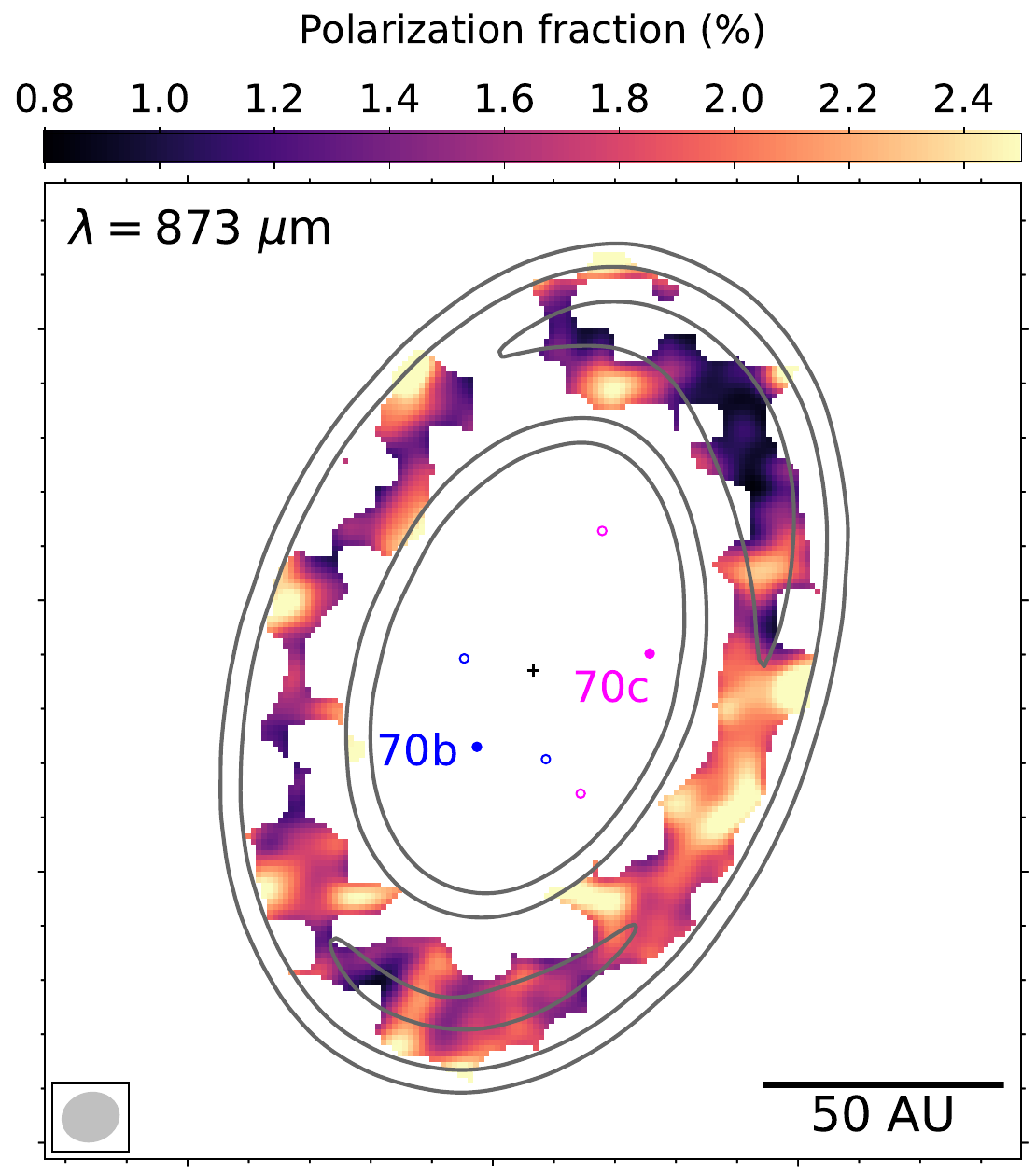} \\
    \end{tabular}
    \end{center}
    \vspace{-0.35cm}
    \caption{
    The 873 $\mu$m full polarization images on PDS\,70.
    Left panel shows the total intensity (color scale) and polarization position angles (line segments). Right panel shows the total intensity (contours) and the polarization fraction.
    Contour levels are 197 $\mu$Jy\,beam$^{-1}$ (5-$\sigma$) $\times$[3, 6, 12]. The cross symbol marks the location of the host star. Filled and open circles mark the locations of PDS\,70b and 70c, and their L4 and L5 points.  
    Synthesized beam is shown in the lower left.
    }
    \label{fig:b7pol}
\end{figure*}

To observationally test the capability of dust grains growing to millimeter sizes in regions with various grain compositions, we can examine the maximum dust grain size ($a_{\mbox{\scriptsize max}}$) by probing the wavelength-dependent continuum spectral indices ($\alpha$; c.f. Appendix \ref{apdx:spid} for definition; \citealt{Beckwith1991ApJ...381..250B,Testi2014prpl.conf..339T}). 
This is because larger dust grains can emit more efficiently at longer wavelength.
In theory, the most reliable constraint on $a_{\mbox{\scriptsize max}}$ can be given when the observations cover wavelengths that are comparable to and larger than $a_{\mbox{\scriptsize max}}$.
However, the values of $a_{\mbox{\scriptsize max}}$ inferred from such analyses often suffer from low precision and may be degenerate.
Observationally, whether or not $a_{\rm max}$ exceeds $\gtrsim$10$^{3}$ $\mu$m on 10$^{0}$--10$^{2}$ AU scales has been a long-standing issue \citep{Testi2014prpl.conf..339T,Guidi2022A&A...664A.137G,Sierra2025MNRAS.541.3101S}.

Investigation into the effects of dust self-scattering shed light on this problem.
It was realized that the spectral index can be  lowered in a narrow wavelength range close to $\lambda=2\pi a_{\rm max}$ in which the dust albedo is high and is decreasing with wavelength.
Confirming such lowering of spectral indices in multi-wavelength observations (e.g., based on fitting spectral energy distributions) therefore can precisely constrain $a_{\mbox{\scriptsize max}}$ to $\lesssim\lambda/2\pi$ \citep{Liu2019ApJ...877L..22L}.
For example, when dust emission is optically thick and falls within the Rayleigh-Jeans limit, resolving $\alpha\lesssim2$ at 1200 $\mu$m wavelength implies $a_{\rm max}\lesssim$200 $\mu$m.

In addition, self-scattered dust emission in an inclined dusty disk is linearly polarized \citep{Kataoka2015ApJ...809...78K}.
The polarization position angles preferentially align with the projected minor axis in an inclined disk, while they preferentially align in the radial directions in geometrically narrow, optically thin dusty rings \citep{Kataoka2015ApJ...809...78K,Pohl2016A&A...593A..12P,Yang2016MNRAS.456.2794Y}.
Such features are most prominent when $a_{\rm max}$ is close to the Mie region $a_{\rm max}\sim\lambda/2\pi$ \citep{Kataoka2015ApJ...809...78K,Yang2016MNRAS.456.2794Y}.
Theoretically, for spatially resolved observations at a given wavelength $\lambda$, the polarization percentage becomes much higher at the locations where $a_{\mbox{\scriptsize max}}$ is very close to $\lambda/2\pi$ (e.g., 1--3\%; \citealt{Kataoka2015ApJ...809...78K}).
Comparing polarization observations with resolved total intensity spectral modeling is presently the most effective technique for accurately determining $a_{\text{max}}$ within the range of several tens of micrometers to centimeter sizes.

Direct near-infrared imaging observations have resolved two massive planets in the PDS\,70 system ($d\sim$112.4 pc; \citealt{GAIA_2023A&A...674A...1G}), namely PDS\,70b and 70c.
The masses of these two planets were estimated to be 3.2$^{+0.3}_{-1.6}$ $M_{\mbox{\scriptsize Jup}}$ and $\lesssim$7.5--13.6 $M_{\mbox{\scriptsize Jup}}$, respectively \citep{Keppler2018A&A...617A..44K,Haffert2019NatAs...3..749H,Wang2021AJ....161..148W}.
The host star, PDS\,70, is a K7-type (0.85 $M_{\odot}$; \citealt{Keppler2019A&A...625A.118K}), $\sim$5\,Myr weak-line T Tauri type star \citep{Keppler2018A&A...617A..44K,Muller2018A&A...617L...2M}.
The previous near-infrared imaging observations \citep{Hashimoto2012ApJ...758L..19H,Keppler2018A&A...617A..44K,Keppler2019A&A...625A.118K} resolved that PDS\,70 is associated with an extended disk of $\sim$100\,AU radius, which presents a gap extending from $\sim$17--60 AU radii.
The high angular resolution (sub)millimeter imaging observations \citep{Hashimoto2015ApJ...799...43H,Isella2019ApJ...879L..25I,Keppler2019A&A...625A.118K,Facchini2021AJ....162...99F,Casassus2022MNRAS.513.5790C,Liu2024ApJ...972..163L,Doi2024ApJ...974L..25D,Sierra2025MNRAS.541.3101S,Dominguez-Jamett2025arXiv250721970D} have resolved a $\sim$55--100 AU dusty ring and an inner disk at $<$17 AU radius. 
Constraining $a_{\mbox{\scriptsize max}}$ in this system provides the most direct indication for the natal environment of these two planets. 

We performed full polarization observations toward PDS\,70, at an angular resolution of $\lesssim$0$''$.1, and at a wavelength of $\sim$873 $\mu$m.
These are the first (sub)millimeter polarization observations of a disk hosting protoplanets that have been confirmed by direct imaging. 
In addition, we performed the complementary, $\sim$0$''$.1 resolution observations of total intensities (i.e., Stokes {\it I}) at $\sim$2068 and $\sim$3075 $\mu$m wavelengths, and compared them with the publicly available observations at $\sim$1226 and 1287 $\mu$m wavelengths at similar angular resolutions.
We provide an outline of these observations in Section \ref{sec:data}.
The observational results are presented in Section \ref{sec:results}.
Our interpretation for the observational results and the physical implications are discussed in Section \ref{sec:discussion}.
The conclusion is given in Section \ref{sec:conclusion}.
Appendix \ref{apdx:data} provides the details of data calibration and imaging. 
Appendix \ref{apdx:spid} described how we measured spectral indices.
While the present study has an emphasis on presenting the polarization observational results, nevertheless, we have produced a working model to compare with the observations using Monte Carlo radiative transfer simulations.
We outline how the model was produced in Appendix \ref{apdx:radmc}.
More sophisticated modeling is beyond the scope of the present paper, which will be investigated in near future.
Appendix \ref{apdx:vfrag} introduces how the dust fragmentation velocity ($v_{\rm frag}$) can be estimated.

\begin{figure*}
    \hspace{-0.5cm}
    \begin{tabular}{ p{4.2cm} p{4.2cm} p{4.2cm} p{4.2cm} }
    \includegraphics[height=5.3cm]{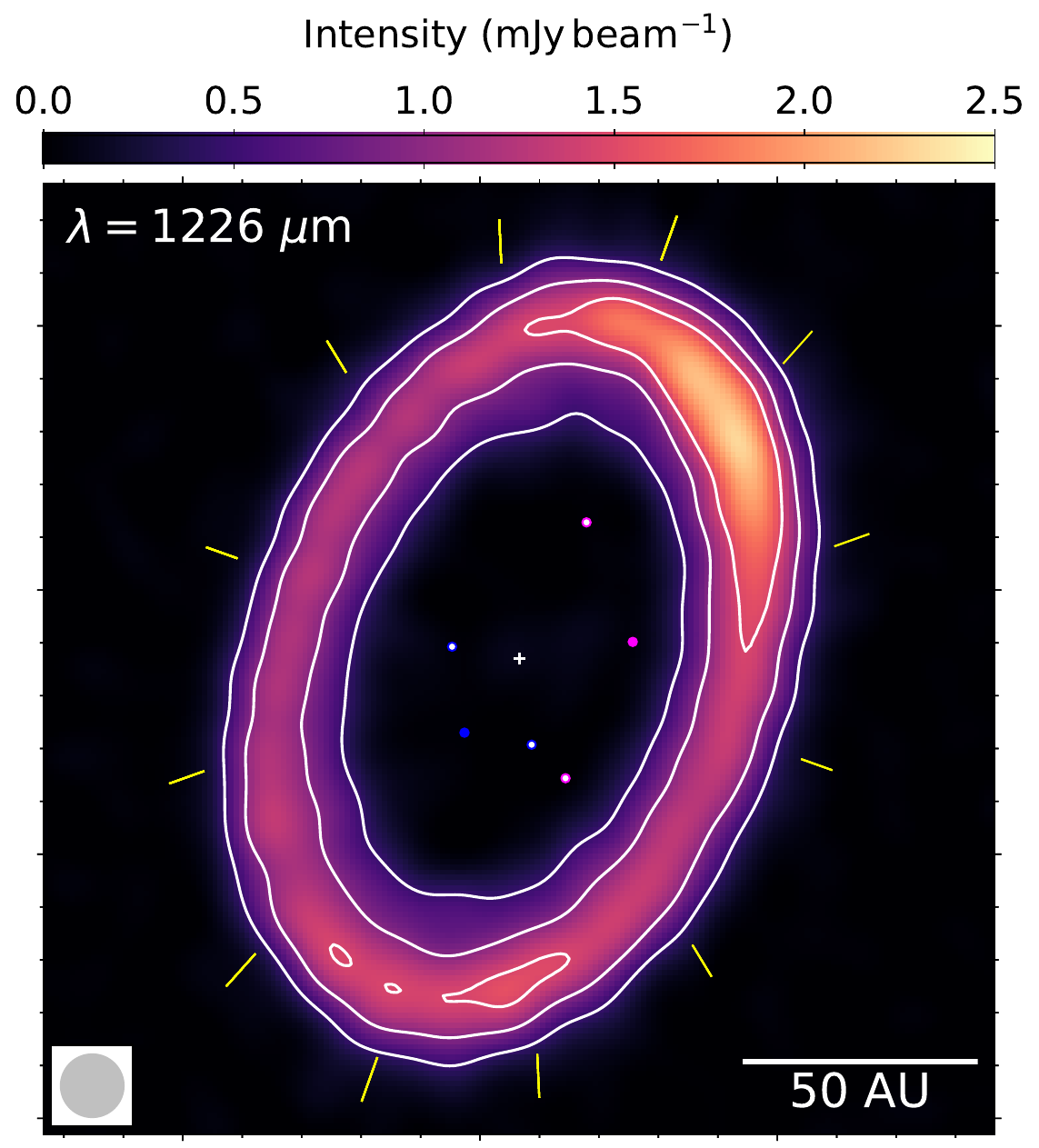} &
    \includegraphics[height=5.3cm]{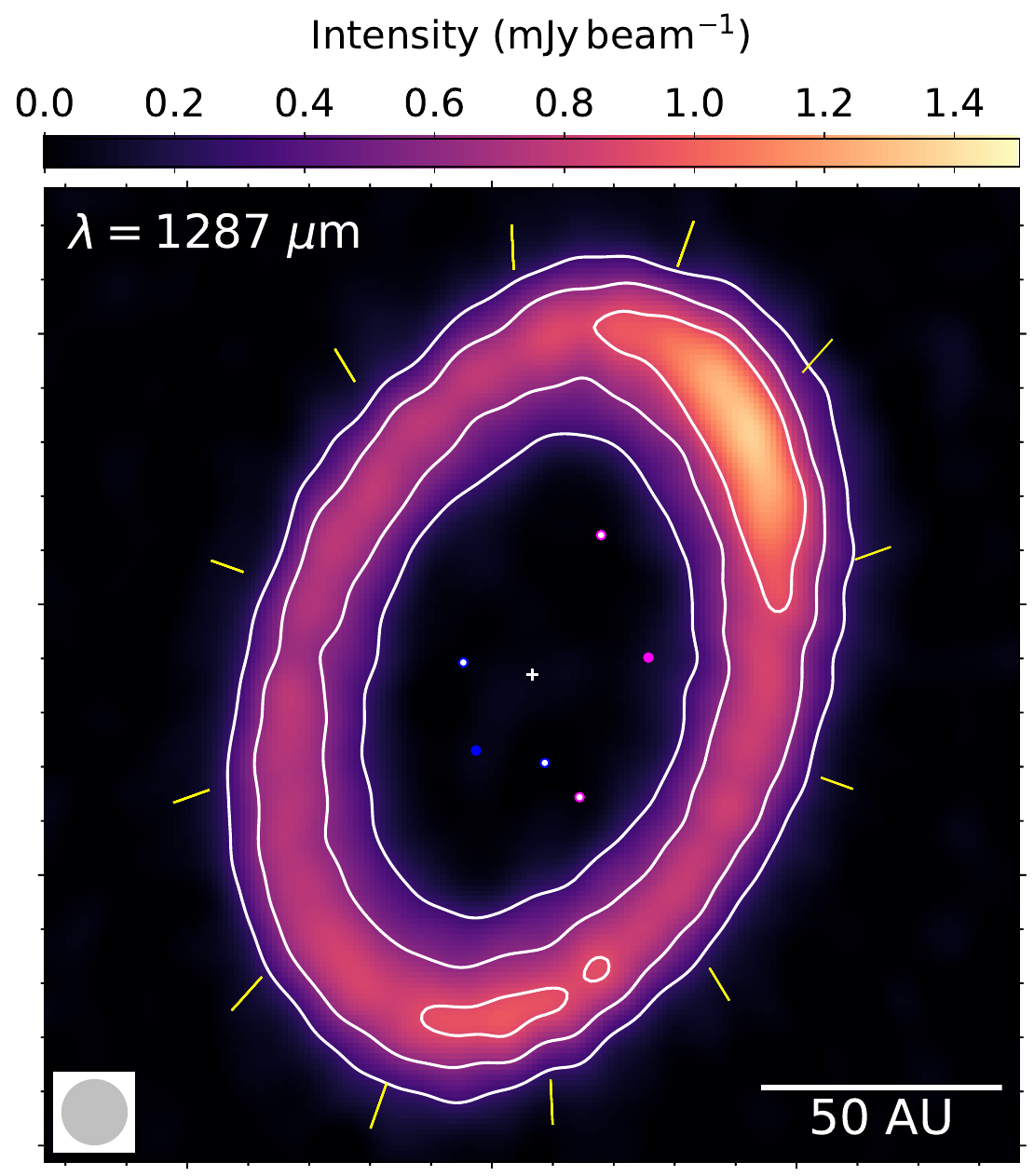} &
    \includegraphics[height=5.3cm]{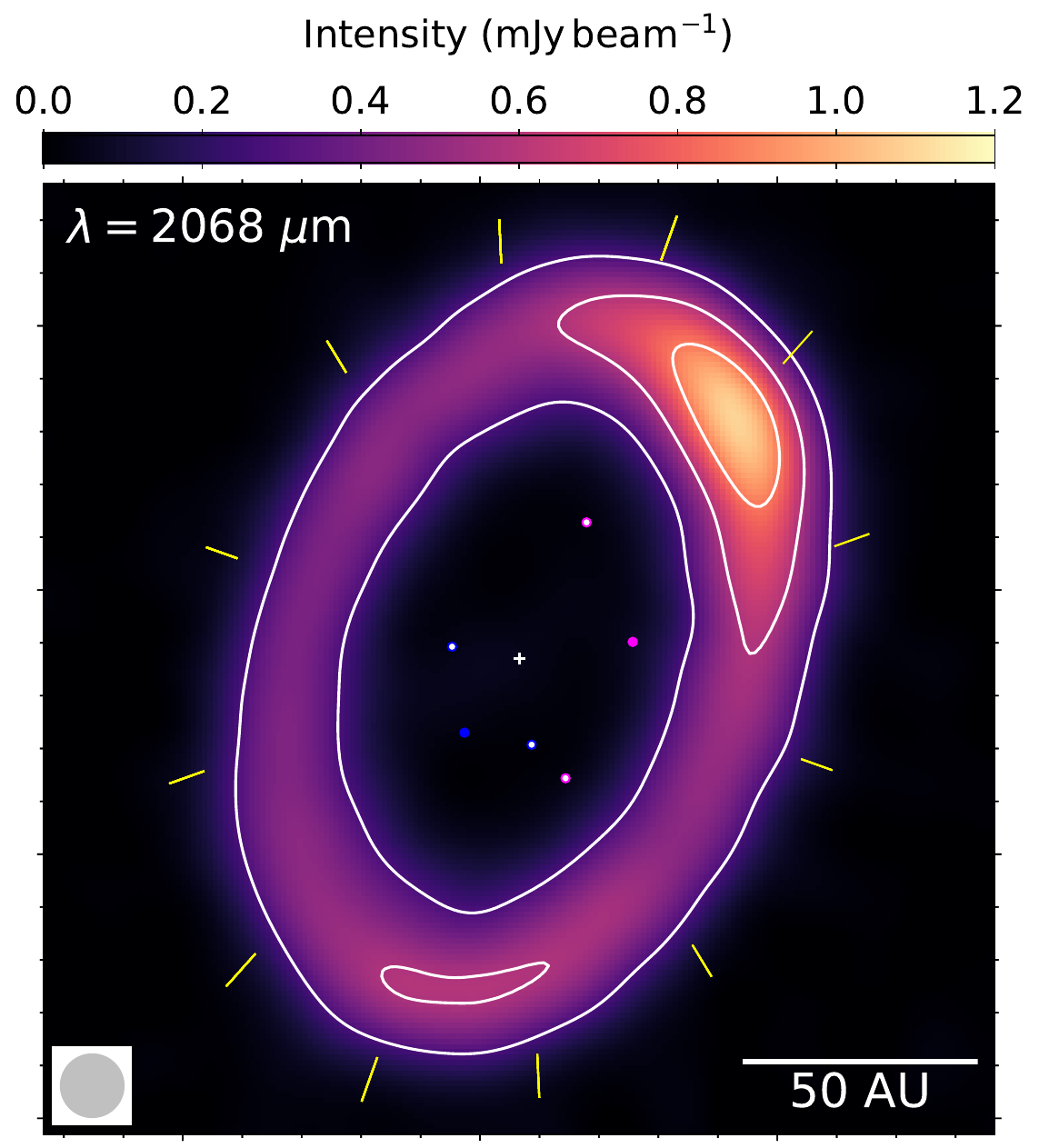} &
    \includegraphics[height=5.3cm]{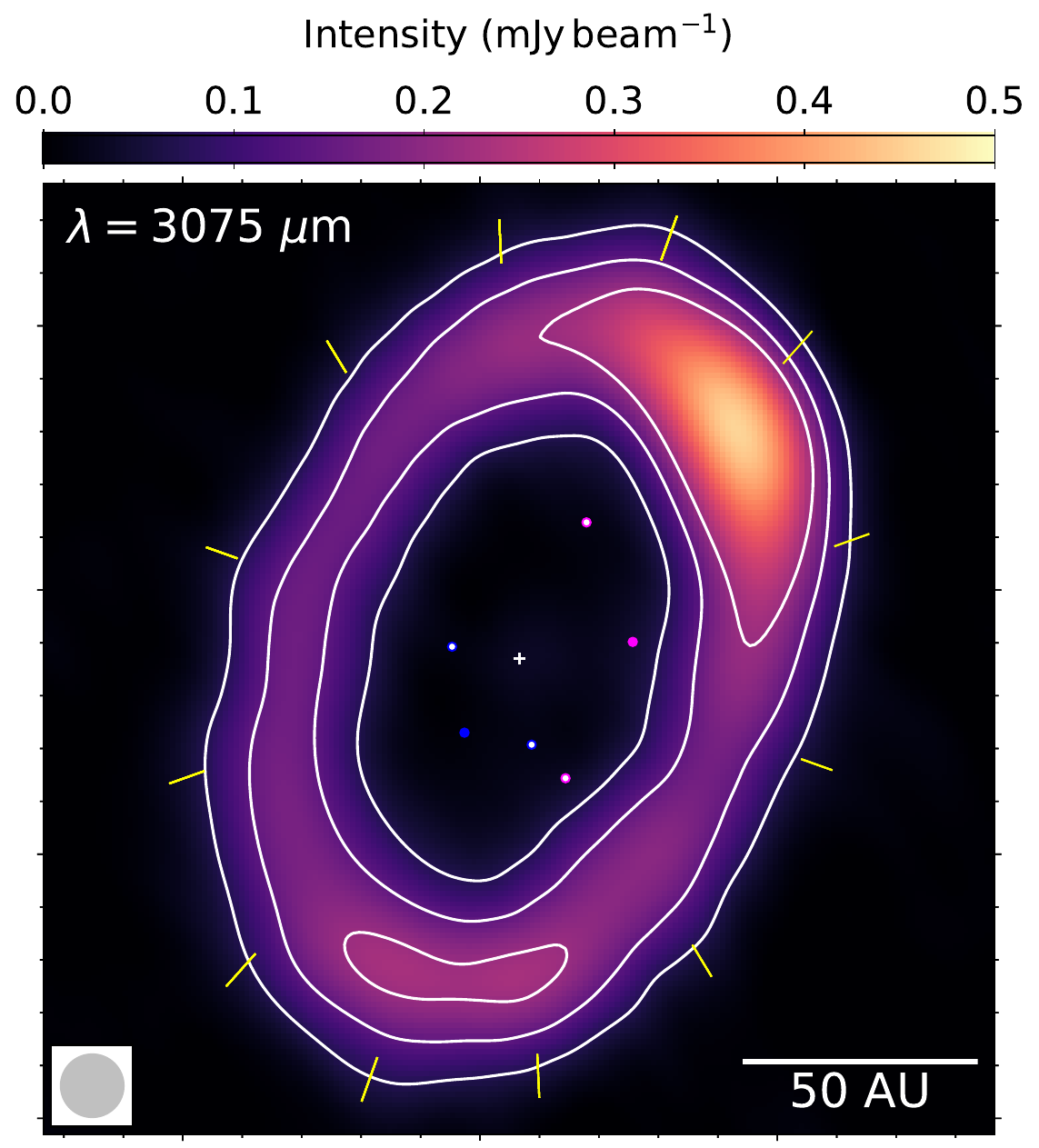} \\
    \end{tabular}
    \vspace{-0.35cm}
    \caption{Total intensity images at 1226, 1287, 2068, and 3075 $\mu$m wavelengths.
    Contours for the $\lambda=$1226 $\mu$m image are 81 $\mu$Jy\,beam$^{-1}$  (3-$\sigma$) $\times$[6, 12, 18]; contours for the $\lambda=$1287 $\mu$m image are 60 $\mu$Jy\,beam$^{-1}$  (3-$\sigma$) $\times$[5, 10, 15]; contours for the $\lambda=$2068 $\mu$m image are 36 $\mu$Jy\,beam$^{-1}$  (3-$\sigma$) $\times$[8,16,24]; contours for the $\lambda=$3075 $\mu$m image are 18 $\mu$Jy\,beam$^{-1}$  (3-$\sigma$) $\times$[4,8,12].
    Symbols are the same as those in Figure \ref{fig:b7pol}.
    Synthesized beams are shown in the lower left. 
    Yellow line segments indicate the deprojected position angles starting from 0$^{\circ}$ (defined at the major axis in the northwest), with 30$^{\circ}$ intervals. 
    }
    \label{fig:maps}
\end{figure*}

\section{Data} \label{sec:data}

\subsection{New ALMA Observations}

We resolved the (sub)millimeter dust continuum emission using the Atacama Large Millimeter/Submillimeter Array (ALMA).
We carried out full polarization observations at Band 7, and carried out total intensity (Stokes {\it I}) observations at Bands 3 and 4.
Basic information of these observations is summarized in Table \ref{tab:obs}.
More technical details are given as follows. 

\subsubsection{Band 7 full polarization observations}

There were 9 executions of high angular resolution observations, which were distributed on seven dates.
Each execution has a $\sim$3 hours' duration.
There was 1 additional execution of short-spacing observations, which was in a relatively compact array configuration to recover the spatially extended emission.
Concatenating all Band 7 observations yielded a 16--4002 $k\lambda$ coverage of {\it uv} distances, which provides a maximum recoverable angular scale (MRS; estimated by $0.6\lambda/D$, where $D$ is the shortest {\it uv} distance) of 3$''$.9.
This MRS is considerably larger than the projected major axis of the target source, PDS~70 (\citealt{Hashimoto2015ApJ...799...43H,Long2018ApJ...858..112L}).
Therefore, our Band 7 observations are not subject to missing fluxes. 

The correlator was configured to simultaneously provide full polarization products (XX, XY, YX, YY) in the four spectral windows that centered at 336.5, 338.5, 348.5, and 350.5 GHz frequencies.
Each spectral window has a 1.875 GHz bandwidth, which was sampled by 1920 spectral channels.
The mean wavelength of these observations was 873 $\mu$m.

Besides our target source, PDS\,70, the observations in each execution covered the absolute flux, passband, complex gain, and polarization calibrators. 
The calibrators for each epoch of observations were automatically selected by the observing script generator during the Phase 2 Group (P2G) stage.

\subsubsection{Band 3 and Band 4 dual polarization observations} 

There were 2 executions of high angular resolution observations and 1 execution of short-spacing observations at Band 3 (Table \ref{tab:obs}). 
Concatenating all Band 3 observations yielded a 4.2--2979 $k\lambda$ coverage of {\it uv} distances, which corresponds to an MRS of 14$''$.7.
There were 1 executions of high angular resolution observations and 1 execution of short-spacing observations at Band 3 (Table \ref{tab:obs}). 
Concatenating all Band 4 observations yielded a 6.5--3919 $k\lambda$ coverage of {\it uv} distances, which corresponds to an MRS of 9$''$.5.

For the Band 3 observations, the correlator was configured to provide dual polarization products (XX, YY) in the four spectral windows that centered at 90.5, 92.5, 102.5, 104.5 GHz; for the Band 4 observations, the correlator was configured to provide dual polarization products in the four spectral windows that centered at 138.0, 140.0, 150.0, 152.0 GHz.
Each spectral window has a 1.875 GHz bandwidth, which was sampled by 3840 spectral channels.
The mean wavelengths of these Band 3 and Band 4 observations were 3075 $\mu$mm and 2068 $\mu$m, respectively.

The observations in each execution of observations covered the target source PDS\,70 and the absolute flux, passband, and complex gain calibrators, which can be looked up at the ALMA data archive\footnote{\url{https://almascience.nrao.edu/aq/}}.

\subsection{Archival ALMA Band 6 observation}
 
We retrieved the archival ALMA Band 6 observations, which were taken via project 2019.1.01619.S.
There were two independent spectral tunings. 
Each spectral tuning employed one broad bandwidth (1.875 GHz) spectral window for observing continuum emission at high signal-to-noise ratios (SNR), and several narrow bandwidth spectral windows for observing spectral lines. 
We only utilized the data taken from the broad bandwidth spectral windows, which have centroid wavelengths of 1226 $\mu$m and 1287 $\mu$m, respectively.

The observations at each tuning included short-spacing observations, which provided $\sim$0$''$.4 angular resolutions, and high angular resolution observations, which provided $\sim$0$''$.1 angular resolutions. 
Concatenating all 1226 $\mu$m observations yielded a 10.1--2674 $k\lambda$  {\it uv} coverage, while concatenating all 1287 $\mu$m observations yielded a 9.8--2552 $k\lambda$ {\it uv} coverage.
The MRS of these observations are larger than those of our Band 7 observations (Table \ref{tab:obs}), which avoided missing fluxes. 
We refer to the previous publication of these observations \citep{Facchini2021AJ....162...99F} for more technical details. 


\begin{figure*}
    \begin{center}
    \begin{tabular}{ cc }
    \raisebox{0.3cm}{\includegraphics[height=8.5cm]{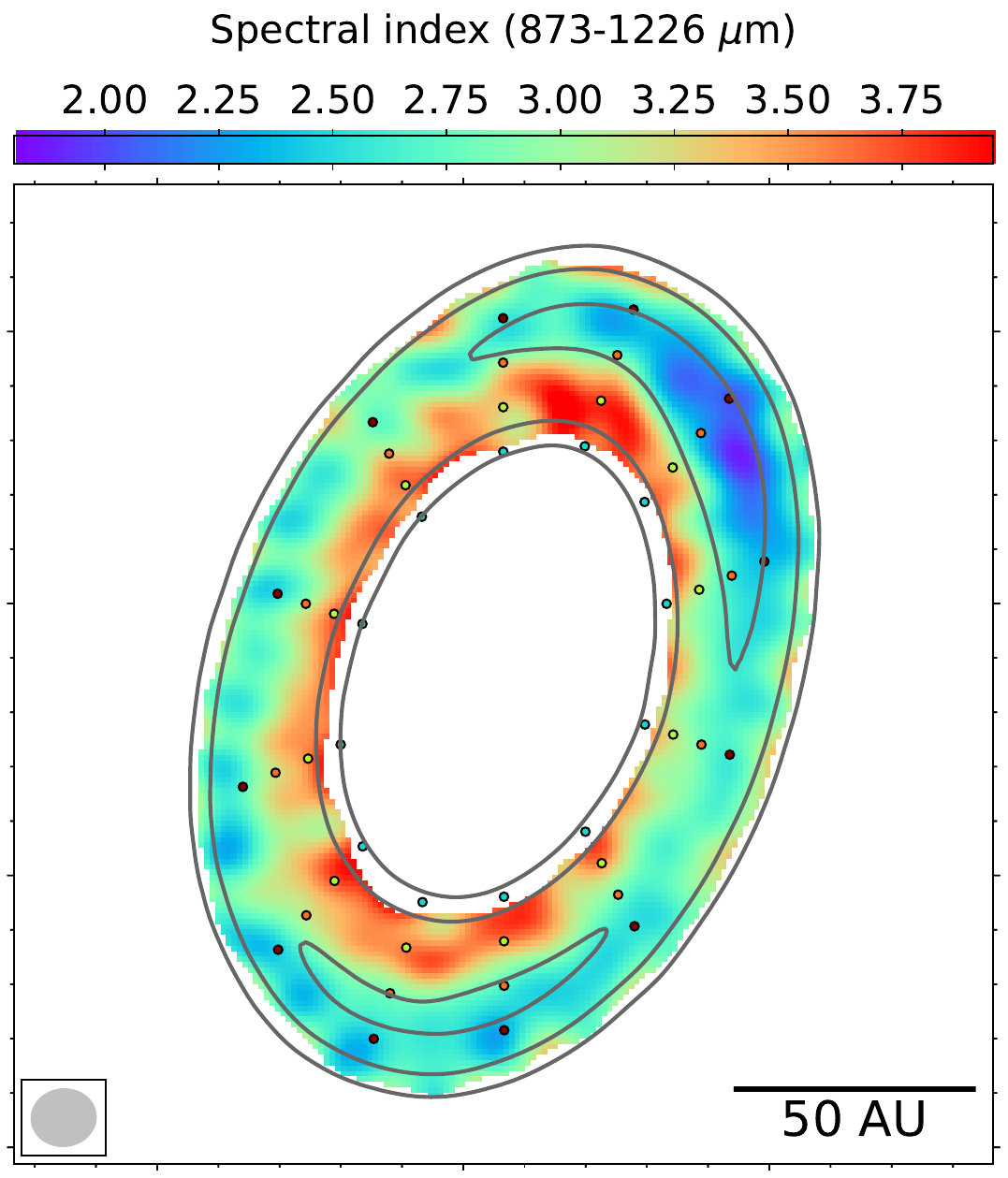}} &
    \includegraphics[height=8.5cm]{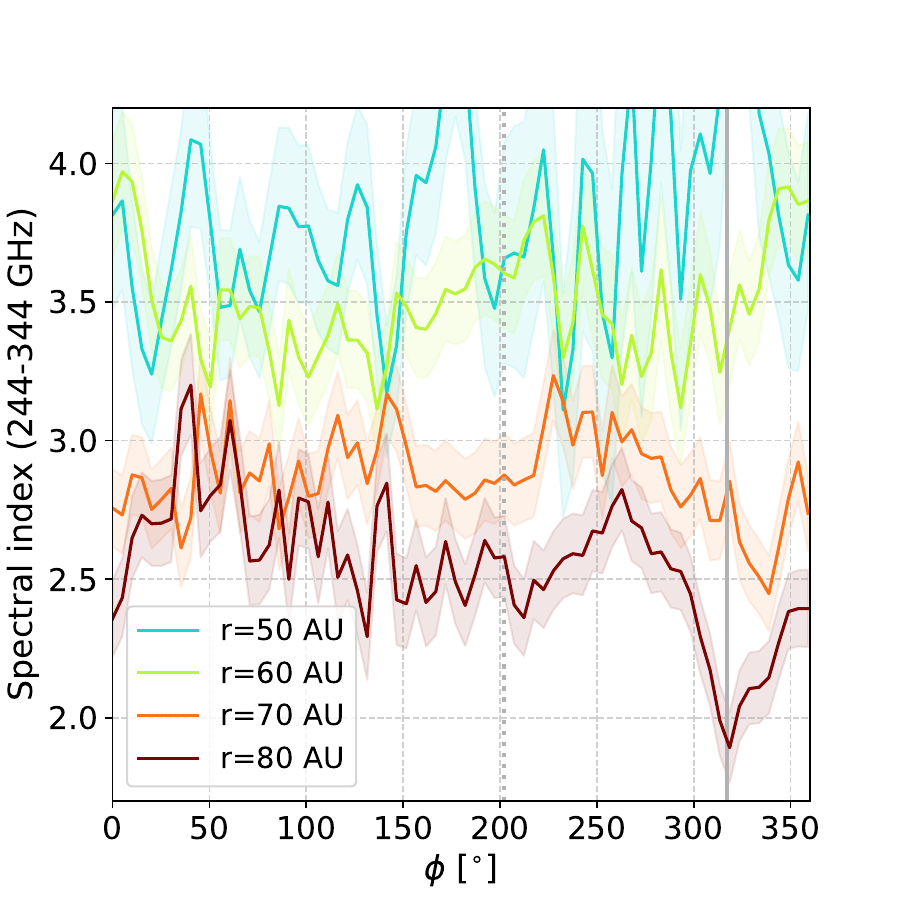} \\
    \end{tabular}
    \end{center}
    \vspace{-0.75cm}
    \caption{
    The 873--1226 $\mu$m spectral index ($\alpha_{\rm b6-b7}$).
    Color image in the left panel shows the spatially resolved distribution of $\alpha_{\rm b6-b7}$. Contours are the same as those in the right panel of Figure \ref{fig:b7pol}. The circles (with 30$^{\circ}$ intervals in deprojected position angles) indicate the 50, 60, 70, and 80 AU deprojected radii. Right panel shows the values of $\alpha_{\rm b6-b7}$ at 50, 60, 70, and 70 AU deprojected radii as functions of deprojected position angle $\phi$. 
    }
    \label{fig:spid}
\end{figure*}

\section{Results} \label{sec:results}
Figure \ref{fig:b7pol} shows the Stokes {\it I} intensities, polarization position angles, and polarization fractions resolved at 873 $\mu$m wavelength.
The Stokes {\it I} intensities at 1226, 1287, 2068, and 3075 $\mu$m wavelengths are  presented in Figure \ref{fig:maps}, which have been smoothed to circular beams to allow diagnosing sub-structures without being confused by the smearing effect of asymmetric beams.

At these wavelengths, the total intensities trace an inclined ring and two crescents situated in the northwest and southwest of the PDS\,70 ring (Figure \ref{fig:maps}).
The higher intensity crescent in the northwest of the ring (hereafter the northwest-crescent) has been reported in the recent studies \citep{Isella2019ApJ...879L..25I,Keppler2019A&A...625A.118K,Liu2024ApJ...972..163L,Doi2024ApJ...974L..25D,Sierra2025MNRAS.541.3101S,Dominguez-Jamett2025arXiv250721970D}.
The fainter one in the southwest (hereafter the southwest crescent) shows consistent angular offset from the projected major axis of the PDS\,70 ring in the multi-wavelength observations (Figure \ref{fig:maps}).
It is more evident in the images at 1226, 1287, 2068, and 3075 $\mu$m. 
The southwest crescent is less prominent at 873 $\mu$m and shorter wavelengths (Figure \ref{fig:b7pol}; \citealt{Isella2019ApJ...879L..25I,Keppler2019A&A...625A.118K,Sierra2025MNRAS.541.3101S,Dominguez-Jamett2025arXiv250721970D}) in spite of the better angular resolutions of these short wavelength observations, due to the high dust optical depths at short wavelengths (more below; \citealt{Liu2024ApJ...972..163L,Sierra2025MNRAS.541.3101S}).
Using Monte Carlo Radiative Transfer (MCRT) modeling (more below; c.f. Appendix \ref{apdx:radmc}), we determined the deprojected centroid position angles (defined north-to-east with respect to the major axis) of northwest-crescent and southwest-crescent to be 317$^{\circ}\pm1^{\circ}$ and 225$^{\circ}\pm3^{\circ}$ respectively.

The morphology of the bulk of the PDS\,70 ring can be described by a geometrically narrow belt (hereafter narrow-ring) centered at a radius of $\sim$77\,AU and superimposed on a fainter, radially more extended ring (hereafter wide-ring).
The wide-ring has a larger extension inward of the 77 AU radius \citep{Isella2019ApJ...879L..25I,Keppler2019A&A...625A.118K,Doi2024ApJ...974L..25D}.  

We detected 873 $\mu$m linear polarization close to the radial range of the narrow-ring. 
The observations resolved azimuthal variations of the polarization fraction in the range of $\sim$1\%--2.5\%. 
The polarization fraction is maximized in the west and presents a minimum at the northwest-crescent (Figure \ref{fig:b7pol}).
The mean polarization position angles is 68$^{\circ}$, which is close to that of the projected minor axis of the PDS\,70 ring.
The polarization position angles are rather uniform, with a modest observed dispersion of 27$^{\circ}$.
The actual dispersion of polarization position angles is smaller, considering that the observations are subject to thermal noise. 

Figure~\ref{fig:spid} shows the spatially resolved spectral indices ($\alpha_{\rm b6-b7}$)  measured at 873–1226 $\mu$m wavelengths.
In general, it is close to the interstellar value, 3.7, at the faint inner rim of the wide-ring, which is consistent with $a_{\rm max}\lesssim$100 $\mu$m \citep{Testi2014prpl.conf..339T}.
The value of $\alpha_{\rm b6-b7}$ is $\sim$2--3 close to the narrow-ring (c.f. \citealt{Fasano2025A&A...699A.373F}).
In addition, $\alpha_{\rm b6-b7}$ shows significant azimuthal variation in the narrow-ring. 
It has a minimum value of 1.86$\pm$0.13 around the northwest crescent,
indicating that dust emission at 873--1226 $\mu$m is likely very optically thick at this location.

\section{Discussion} \label{sec:discussion}

\begin{figure*}
    \begin{center}
     \begin{tabular}{ c c }
     \raisebox{0.5cm}{\includegraphics[height=8.5cm]{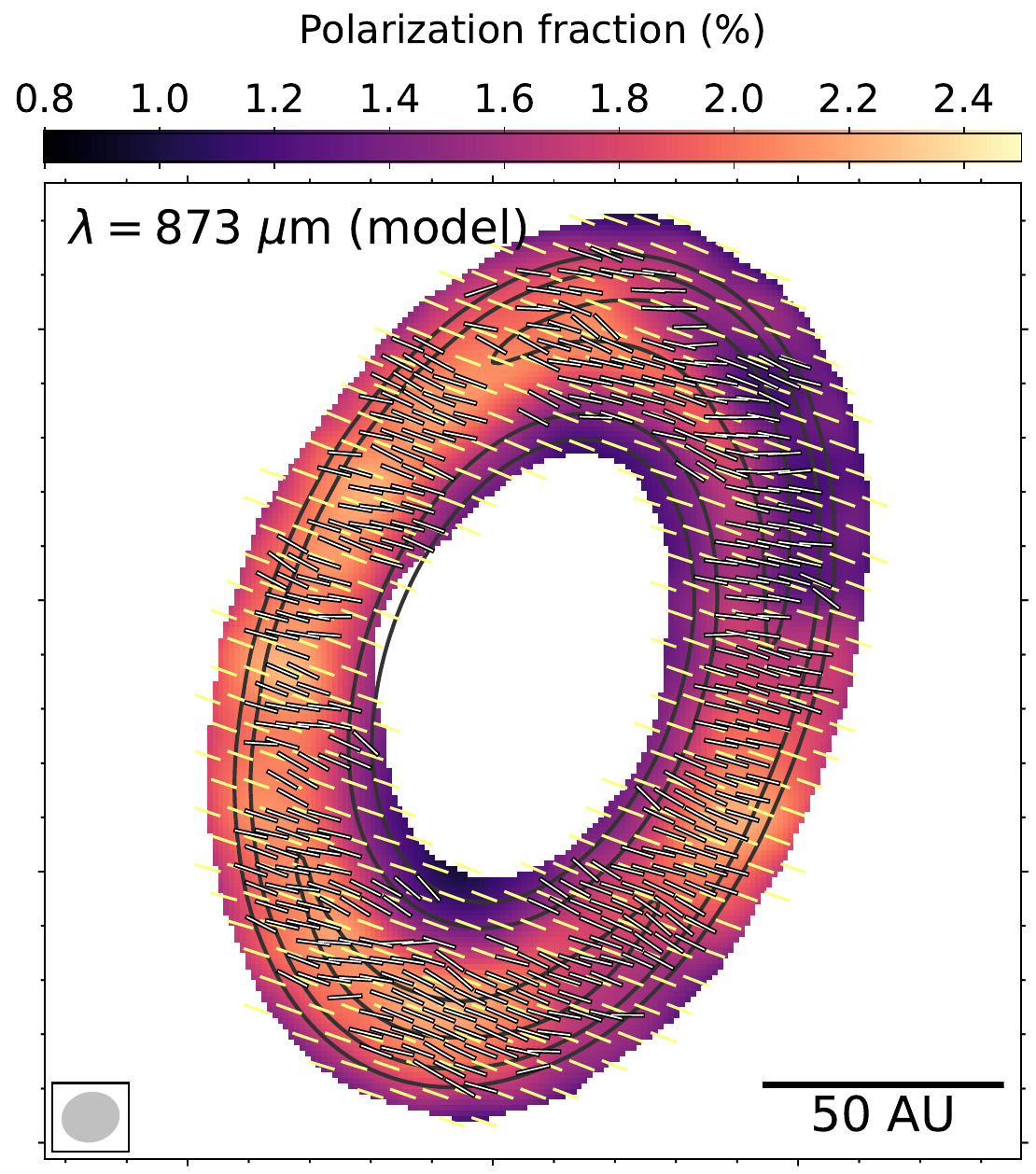}} &
     \includegraphics[width=8cm]{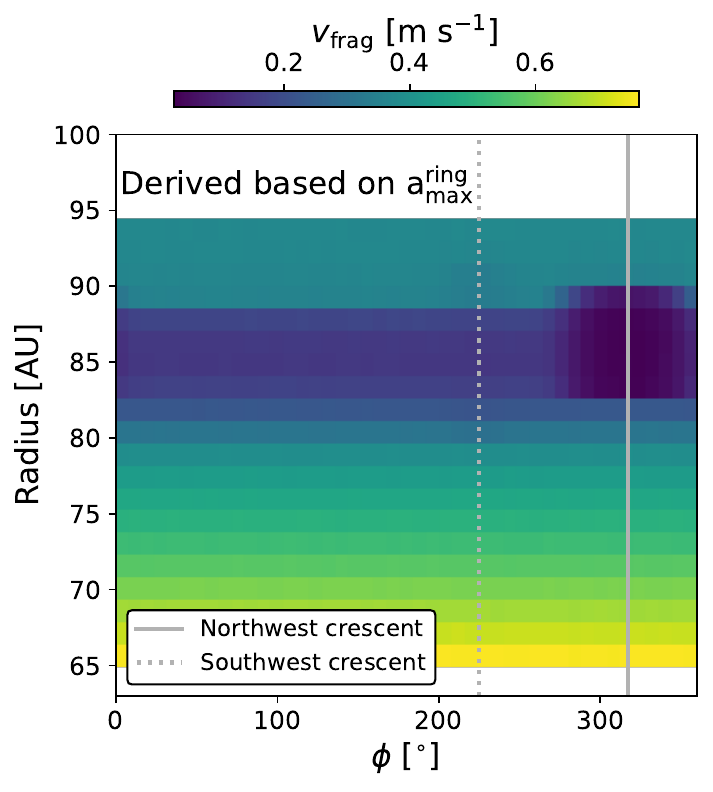} \\
     \end{tabular}
    \end{center}
    \vspace{-0.75cm}
    \caption{Physical properties derived from the fiducial radiative transfer model.
    Left panel shows the synthesized, $\lambda=$873 $\mu$m images of total intensity (contours), polarization fraction (color), and polarization position angle (yellow line segments).
    We overplot the polarization position angles detected in actual observations (Figure \ref{fig:b7pol}).
    The angular resolution of these synthesized images are the same as those in Figure \ref{fig:b7pol}.
    Contour levels are 197 $\mu$Jy\,beam$^{-1}$ (5-$\sigma$) $\times$[3, 6, 12].
    Synthesized beam is shown in the lower left.
    Right panel shows the fragmentation velocity ($v_{\rm frag}$) derived based on the maximum dust grain sizes ($a_{\rm max}$) in the PDS\,70 rings.
    }
    \label{fig:best-fit}
\end{figure*}

The observed 873 $\mu$m polarization and the $\alpha_{\rm b6-b7}$ distributions can be concordantly explained if $a_{\rm max}$ in the bulk of the (sub)millimeter ring is $\lesssim$100 $\mu$m.
This is the first observational indication that the (sub)millimeter luminosity in a protoplantary disk is dominated by $\lesssim$100 $\mu$m sized dust even after the formation of planets. 
The observed polarization position angles appear more like the case of an inclined disk rather than a ring for two reasons.
First, the radial extension of the PDS\,70 submillimeter ring is wide (e.g,. compared to scale height, c.f. Table \ref{tab:bestfit}).
Second, at 873 $\mu$m wavelength, the ring is likely optically thick in the azimuthal direction.
These make the local radiative field around the narrow-ring effectively more like that in a disk.

To help understand the observational results, we constructed a model using Monte Carlo radiative transfer (MCRT) simulations and determined the free parameters in the model using the Markov Chain Monte Carlo (MCMC) method (see Appendix \ref{apdx:radmc}). 
The main purpose of this model is to verify the qualitative interpretation of the observed 873 $\mu$m total intensity and polarization, rather than fitting the intensity distributions in detail. 
Figure~\ref{fig:best-fit} shows the synthesized 873 $\mu$m observations produced from our fiducial model.
The total intensity distribution resembles the actual observations (Figure \ref{fig:b7pol}). 
It reproduces the azimuthally varying, $\sim$1\%--2\% polarization produced due to dust self-scattering, with polarization position angles aligned with the minor axis of the (sub)millimeter ring. 

In this model, the peak dust mass surface density is 0.4 g\,cm$^{-2}$ in the narrow-ring, which are marginally optically thick at $\lesssim$1200 $\mu$m wavelengths.
This results in $\alpha_{\rm b6-b7}\gtrsim$2 in most of the narrow-ring.
The maximum dust mass surface density is 0.093 g\,cm$^{-2}$ in the wide-ring such that the (sub)millimeter ring becomes optically thin at its inner rim. 
The peak dust mass surface density in the northwest-crescent is 11 g\,cm$^{-2}$, which is optically thick at $\lesssim$1200 $\mu$m wavelengths. 
The values of $a_{\rm max}$ are 87 $\mu$m and 79 $\mu$m in the ring and in the northwest-crescent, respectively.
The 873 $\mu$m polarization percentage is suppressed at the northwest-crescent primarily due to high optical depth rather than the spatial variation of $a_{\rm max}$.
The overall dust masses in the azimuthally symmetric rings, the northwest-crescent, and the southest crescent are 103 $M_{\oplus}$, 174 $M_{\oplus}$, and 3.6 $M_{\oplus}$, respectively.
Our present model was based on the assumption of relatively compact dust grains (Appendix \ref{subsub:smalldustdisk}, \ref{subsub:ring}, \ref{subsub:crescent}).
Considering a slight porosity (e.g. $\sim$80\% of volume is vacuum) may make the derived $a_{\rm max}$ values larger by a few times \citep{Ueda2024NatAs...8.1148U}, which needs to be testified by future, multi-wavelength polarization observations and is beyond the scope of this paper.  

The stickiness of dust can be characterized with the fragmentation velocity ($v_{\rm frag}$), which is the largest relative velocity between two colliding dust particles that can still yield coagulation instead of fragmentation.
Assuming that dust growth is limited by fragmentation rather than inward migration, which is realistic in the studies of dusty rings and crescent, $v_{\rm frag}$ can be inferred from the observed $a_{\rm max}$ values (Appendix \ref{apdx:vfrag}; \citealt{Birnstiel2016SSRv..205...41B}).
We found that $v_{\rm frag}$ in the bulk of the (sub)millimeter ring may be a fraction of 1 m\,s$^{-1}$ (Figure \ref{fig:best-fit}). 
Even if we consider a realistic, small dust porosity \citep{Ueda2024NatAs...8.1148U}, $v_{\rm frag}$ (in m\,s$^{-1}$ units) can only be as high as order of unity.
Since the (sub)millimter ring in PDS\,70 is outside of the water snowline, the derived $v_{\rm frag}$ is in line with the recent laboratory experiments and theoretical studies which suggested water-ice-coated dust is not sticky \citep{Musiolik2019ApJ...873...58M,Kimura2020MNRAS.498.1801K}.

A caveat is that the present observations do not rule out the possibility that the observed 873 $\mu$m is partly contributed by emission of aligned dust (\citealt{Hildebrand2000PASP..112.1215H}), which can be tested by future, multi-frequency full polarization ALMA observations.

\begin{table*}\footnotesize
\centering
\caption{ALMA Observations}\label{tab:obs}
\begin{tabular}{ccccccc}
\hline\hline\noalign{\vspace{5pt}}
Start Time & End Time & $b_{\rm min}$  & $b_{\rm max}$ & {\it uv} Distance & Available Antennae & PWV  \\
(UTC)      & (UTC)    & (meters)       & (meters)      & ($k\lambda$)      &                    & (mm) \\
(1)        & (2)      & (3)            & (4)           & (5)               & (6)                & (7)  \\
\noalign{\vspace{2pt}}
\hline\noalign{\vspace{5pt}}
\multicolumn{7}{c}{ Band 7 (high-resolution) } \\
2023-05-23 00:23:41 & 2023-05-23 03:12:31 & 78 & 3637 & 79--3665 & 44 & 0.7--1.1 \\
2023-06-01 22:14:09 & 2023-06-02 03:49:32 & 27 & 3637 & 64--3957 & 43 & 0.4--0.6 \\
2023-06-03 22:39:50 & 2023-06-04 04:05:06 & 27 & 3637 & 29--3986 & 41 & 0.6--0.7 \\
2023-06-07 22:51:33 & 2023-06-08 01:38:49 & 76 & 3637 & 82--4002 & 43 & 0.5--0.8 \\  
2023-06-08 23:22:52 & 2023-06-09 02:10:31 & 27 & 3637 & 79--3624 & 43 & 0.6--0.8 \\
2023-11-27 10:32:06 & 2023-11-27 13:20:29 & 30 & 3697 & 26--3983 & 45 & 0.6--0.8 \\
2023-11-28 10:40:07 & 2023-11-28 13:31:23 & 30 & 3637 & 27--3992 & 46 & 0.7--1.0 \\ 
\multicolumn{7}{c}{ Band 7 (short-spacing) } \\
2023-01-23 09:23:26 & 2023-01-23 11:50:45 &  15 & 783 & 16--823  & 47 & $\sim$0.4 \\
\multicolumn{7}{c}{ Band 4 (high-resolution) } \\
2023-06-19 23:34:30 & 2023-06-20 00:45:48 & 91 & 8282 & 40--3319 & 44 & 1.8--2.1 \\
\multicolumn{7}{c}{ Band 4 (short-spacing) } \\
2023-03-25 07:26:14 & 2023-03-25 07:41:46 & 15 & 1261 & 6.5--631 & 40 & 3.1--3.2 \\
\multicolumn{7}{c}{ Band 3 (high-resolution) } \\
2023-07-02 21:54:49 & 2023-07-02 23:25:22 & 84 & 8547 & 25--2879 & 42 & 0.7--0.8 \\
2023-07-02 23:46:31 & 2023-07-03 01:17:13 & 84 & 8547 & 24--2979 & 43 & 0.5--0.6 \\
\multicolumn{7}{c}{ Band 3 (short-spacing) } \\
2023-03-26 04:56:48 & 2023-03-26 05:56:13 & 15 & 1397 & 4.2--432 & 46 & 4.7--5.4 \\
\noalign{\vspace{2pt}}\hline
\noalign{\vspace{5pt}}
\end{tabular}

Notes. Columns (1) and (2) are the start and end time of the schedling block. Columns (3) and (4) are the shortest and longest baseline lengths. Column (5) is the range of projected baseline length. Column (6) is the number of available antennae during the observations. Column (7) is the range of precipitable water vapor during the observations.
\end{table*}

\section{Conclusion} \label{sec:conclusion}

We present the ALMA 873 $\mu$m full polarization observations and the complementary Stokes {\it I} observations at 1226, 1287, 2068, and 3075 $\mu$m wavelengths, towards the planet-hosting dusty ring, PDS\,70.
The 873 $\mu$m observations detected $\sim$1\%--2.5\% linear polarization.
The polarization position angles are uniform, which approximately align with the projected minor axis of the PDS\,70 dusty ring. 
Our observational results are consistent with a maximum dust grain size of $\lesssim$100 $\mu$m. 
If this is indeed the case, then our results imply that the fragmentation velocity of the water-ice-coated dust grains may be $\lesssim$1 m\,s$^{-1}$.
Outside of water snowlines, the fragmentation and bouncing barriers may limit dust growth. 


\begin{acknowledgements}
We deeply thank Dr. Conelis Duellemond for the constructive comments.
This paper makes use of the following ALMA data: ADS/JAO.ALMA\#2019.1.01619.S and \#2022.1.01477.S. 
ALMA is a partnership of ESO (representing its member states), NSF (USA) and NINS (Japan), together with NRC (Canada), MOST and ASIAA (Taiwan), and KASI (Republic of Korea), in cooperation with the Republic of Chile. The Joint ALMA Observatory is operated by ESO, AUI/NRAO and NAOJ. 
H.B.L. is supported by the National Science and Technology Council (NSTC) of Taiwan (Grant No. 113-2112-M-110-022-MY3).
PW acknowledges support from the ANID -- Millennium Science Initiative Program -- Center Code NCN2024\_001.
\end{acknowledgements}

%

\bibliographystyle{aa}
\bibliography{main}

@ARTICLE{Beckwith1991ApJ...381..250B,
       author = {{Beckwith}, Steven V.~W. and {Sargent}, Anneila I.},
        title = "{Particle Emissivity in Circumstellar Disks}",
      journal = {\apj},
         year = 1991,
        month = nov,
       volume = {381},
        pages = {250},
          doi = {10.1086/170646},
       adsurl = {https://ui.adsabs.harvard.edu/abs/1991ApJ...381..250B}
}

@ARTICLE{Bi2021ApJ...912..107B,
       author = {{Bi}, Jiaqing and {Lin}, Min-Kai and {Dong}, Ruobing},
        title = "{Puffed-up Edges of Planet-opened Gaps in Protoplanetary Disks. I. Hydrodynamic Simulations}",
      journal = {\apj},
         year = 2021,
        month = may,
       volume = {912},
       number = {2},
          eid = {107},
        pages = {107},
          doi = {10.3847/1538-4357/abef6b},
archivePrefix = {arXiv},
       eprint = {2103.09254},
 primaryClass = {astro-ph.EP},
       adsurl = {https://ui.adsabs.harvard.edu/abs/2021ApJ...912..107B}
}

@ARTICLE{Birnstiel2016SSRv..205...41B,
       author = {{Birnstiel}, T. and {Fang}, M. and {Johansen}, A.},
        title = "{Dust Evolution and the Formation of Planetesimals}",
      journal = {\ssr},
         year = 2016,
        month = dec,
       volume = {205},
       number = {1-4},
        pages = {41-75},
          doi = {10.1007/s11214-016-0256-1},
archivePrefix = {arXiv},
       eprint = {1604.02952},
 primaryClass = {astro-ph.SR},
       adsurl = {https://ui.adsabs.harvard.edu/abs/2016SSRv..205...41B}
}

@ARTICLE{Birnstiel2018ApJ...869L..45B,
       author = {{Birnstiel}, Tilman and {Dullemond}, Cornelis P. and {Zhu}, Zhaohuan and {Andrews}, Sean M. and {Bai}, Xue-Ning and {Wilner}, David J. and {Carpenter}, John M. and {Huang}, Jane and {Isella}, Andrea and {Benisty}, Myriam and {P{\'e}rez}, Laura M. and {Zhang}, Shangjia},
        title = "{The Disk Substructures at High Angular Resolution Project (DSHARP). V. Interpreting ALMA Maps of Protoplanetary Disks in Terms of a Dust Model}",
      journal = {\apjl},
         year = 2018,
        month = dec,
       volume = {869},
       number = {2},
          eid = {L45},
        pages = {L45},
          doi = {10.3847/2041-8213/aaf743},
archivePrefix = {arXiv},
       eprint = {1812.04043},
 primaryClass = {astro-ph.SR},
       adsurl = {https://ui.adsabs.harvard.edu/abs/2018ApJ...869L..45B}
}

@ARTICLE{Cazzoletti2018A&A...619A.161C,
       author = {{Cazzoletti}, P. and {van Dishoeck}, E.~F. and {Pinilla}, P. and {Tazzari}, M. and {Facchini}, S. and {van der Marel}, N. and {Benisty}, M. and {Garufi}, A. and {P{\'e}rez}, L.~M.},
        title = "{Evidence for a massive dust-trapping vortex connected to spirals. Multi-wavelength analysis of the HD 135344B protoplanetary disk}",
      journal = {\aap},
         year = 2018,
        month = nov,
       volume = {619},
          eid = {A161},
        pages = {A161},
          doi = {10.1051/0004-6361/201834006},
archivePrefix = {arXiv},
       eprint = {1809.04160},
 primaryClass = {astro-ph.EP},
       adsurl = {https://ui.adsabs.harvard.edu/abs/2018A&A...619A.161C}
}

@ARTICLE{CASA2022PASP..134k4501C,
       author = {{CASA Team} and {Bean}, Ben and {Bhatnagar}, Sanjay and {Castro}, Sandra and {Donovan Meyer}, Jennifer and {Emonts}, Bjorn and {Garcia}, Enrique and {Garwood}, Robert and {Golap}, Kumar and {Gonzalez Villalba}, Justo and {Harris}, Pamela and {Hayashi}, Yohei and {Hoskins}, Josh and {Hsieh}, Mingyu and {Jagannathan}, Preshanth and {Kawasaki}, Wataru and {Keimpema}, Aard and {Kettenis}, Mark and {Lopez}, Jorge and {Marvil}, Joshua and {Masters}, Joseph and {McNichols}, Andrew and {Mehringer}, David and {Miel}, Renaud and {Moellenbrock}, George and {Montesino}, Federico and {Nakazato}, Takeshi and {Ott}, Juergen and {Petry}, Dirk and {Pokorny}, Martin and {Raba}, Ryan and {Rau}, Urvashi and {Schiebel}, Darrell and {Schweighart}, Neal and {Sekhar}, Srikrishna and {Shimada}, Kazuhiko and {Small}, Des and {Steeb}, Jan-Willem and {Sugimoto}, Kanako and {Suoranta}, Ville and {Tsutsumi}, Takahiro and {van Bemmel}, Ilse M. and {Verkouter}, Marjolein and {Wells}, Akeem and {Xiong}, Wei and {Szomoru}, Arpad and {Griffith}, Morgan and {Glendenning}, Brian and {Kern}, Jeff},
        title = "{CASA, the Common Astronomy Software Applications for Radio Astronomy}",
      journal = {\pasp},
         year = 2022,
        month = nov,
       volume = {134},
       number = {1041},
          eid = {114501},
        pages = {114501},
          doi = {10.1088/1538-3873/ac9642},
archivePrefix = {arXiv},
       eprint = {2210.02276},
 primaryClass = {astro-ph.IM},
       adsurl = {https://ui.adsabs.harvard.edu/abs/2022PASP..134k4501C}
}

@ARTICLE{Casassus2022MNRAS.513.5790C,
       author = {{Casassus}, Simon and {C{\'a}rcamo}, Miguel},
        title = "{Variable structure in the PDS 70 disc and uncertainties in radio-interferometric image restoration}",
      journal = {\mnras},
         year = 2022,
        month = jul,
       volume = {513},
       number = {4},
        pages = {5790-5798},
          doi = {10.1093/mnras/stac1285},
archivePrefix = {arXiv},
       eprint = {2204.08589},
 primaryClass = {astro-ph.EP},
       adsurl = {https://ui.adsabs.harvard.edu/abs/2022MNRAS.513.5790C}
}

@ARTICLE{Chung2024ApJS..273...29C,
       author = {{Chung}, Chia-Ying and {Andrews}, Sean M. and {Gurwell}, Mark A. and {Wright}, Melvyn and {Long}, Feng and {Xu}, Wenrui and {Liu}, Hauyu Baobab},
        title = "{SMA 200{\textendash}400 GHz Survey for Dust Properties in the Icy Class II Disks in the Taurus Molecular Cloud}",
      journal = {\apjs},
         year = 2024,
        month = aug,
       volume = {273},
       number = {2},
          eid = {29},
        pages = {29},
          doi = {10.3847/1538-4365/ad528b},
archivePrefix = {arXiv},
       eprint = {2405.19867},
 primaryClass = {astro-ph.EP},
       adsurl = {https://ui.adsabs.harvard.edu/abs/2024ApJS..273...29C}
}

@ARTICLE{Chung2025ApJS..277...45C,
       author = {{Chung}, Chia-Ying and {Tsai}, An-Li and {Wright}, Melvyn and {Xu}, Wenrui and {Long}, Feng and {Gurwell}, Mark A. and {Liu}, Hauyu Baobab},
        title = "{The 4{\textendash}400 GHz Survey for the 32 Class II Disks in the Taurus Molecular Cloud}",
      journal = {\apjs},
         year = 2025,
        month = apr,
       volume = {277},
       number = {2},
          eid = {45},
        pages = {45},
          doi = {10.3847/1538-4365/adb717},
archivePrefix = {arXiv},
       eprint = {2502.14342},
 primaryClass = {astro-ph.EP},
       adsurl = {https://ui.adsabs.harvard.edu/abs/2025ApJS..277...45C}
}

@ARTICLE{Delussu2024A&A...688A..81D,
       author = {{Delussu}, Luca and {Birnstiel}, Tilman and {Miotello}, Anna and {Pinilla}, Paola and {Rosotti}, Giovanni and {Andrews}, Sean M.},
        title = "{Population synthesis models indicate a need for early and ubiquitous disk substructures}",
      journal = {\aap},
         year = 2024,
        month = aug,
       volume = {688},
          eid = {A81},
        pages = {A81},
          doi = {10.1051/0004-6361/202450328},
archivePrefix = {arXiv},
       eprint = {2405.14501},
 primaryClass = {astro-ph.EP},
       adsurl = {https://ui.adsabs.harvard.edu/abs/2024A&A...688A..81D}
}

@ARTICLE{Doi2024ApJ...974L..25D,
       author = {{Doi}, Kiyoaki and {Kataoka}, Akimasa and {Liu}, Hauyu Baobab and {Yoshida}, Tomohiro C. and {Benisty}, Myriam and {Dong}, Ruobing and {Yamato}, Yoshihide and {Hashimoto}, Jun},
        title = "{Asymmetric Dust Accumulation of the PDS 70 Disk Revealed by ALMA Band 3 Observations}",
      journal = {\apjl},
         year = 2024,
        month = oct,
       volume = {974},
       number = {2},
          eid = {L25},
        pages = {L25},
          doi = {10.3847/2041-8213/ad7f51},
archivePrefix = {arXiv},
       eprint = {2408.09216},
 primaryClass = {astro-ph.EP},
       adsurl = {https://ui.adsabs.harvard.edu/abs/2024ApJ...974L..25D}
}

@ARTICLE{Dominguez-Jamett2025arXiv250721970D,
       author = {{Dominguez-Jamett}, Oriana and {Casassus}, Simon and {Liu}, Hauyu Baobab and {Aoyama}, Yuhiko and {Carcamo}, Miguel and {Weber}, Philipp and {Chrenko}, Ondrej and {Marleau}, Gabriel-Dominique and {Ercolano}, Barbara and {Szulagyi}, Judit},
        title = "{Multi-frequency observations of PDS 70c: Radio emission mechanisms in the circum-planetary environment}",
      journal = {arXiv e-prints},
         year = 2025,
        month = jul,
          eid = {arXiv:2507.21970},
        pages = {arXiv:2507.21970},
          doi = {10.48550/arXiv.2507.21970},
archivePrefix = {arXiv},
       eprint = {2507.21970},
 primaryClass = {astro-ph.EP},
       adsurl = {https://ui.adsabs.harvard.edu/abs/2025arXiv250721970D}
}

@ARTICLE{Dong2012ApJ...760..111D,
       author = {{Dong}, Ruobing and {Hashimoto}, Jun and {Rafikov}, Roman and {Zhu}, Zhaohuan and {Whitney}, Barbara and {Kudo}, Tomoyuki and {Muto}, Takayuki and {Brandt}, Timothy and {McClure}, Melissa K. and {Wisniewski}, John and {Abe}, L. and {Brandner}, W. and {Carson}, J. and {Egner}, S. and {Feldt}, M. and {Goto}, M. and {Grady}, C. and {Guyon}, O. and {Hayano}, Y. and {Hayashi}, M. and {Hayashi}, S. and {Henning}, T. and {Hodapp}, K.~W. and {Ishii}, M. and {Iye}, M. and {Janson}, M. and {Kandori}, R. and {Knapp}, G.~R. and {Kusakabe}, N. and {Kuzuhara}, M. and {Kwon}, J. and {Matsuo}, T. and {McElwain}, M. and {Miyama}, S. and {Morino}, J. -I. and {Moro-Martin}, A. and {Nishimura}, T. and {Pyo}, T. -S. and {Serabyn}, E. and {Suto}, H. and {Suzuki}, R. and {Takami}, M. and {Takato}, N. and {Terada}, H. and {Thalmann}, C. and {Tomono}, D. and {Turner}, E. and {Watanabe}, M. and {Yamada}, T. and {Takami}, H. and {Usuda}, T. and {Tamura}, M.},
        title = "{The Structure of Pre-transitional Protoplanetary Disks. I. Radiative Transfer Modeling of the Disk+Cavity in the PDS 70 System}",
      journal = {\apj},
         year = 2012,
        month = dec,
       volume = {760},
       number = {2},
          eid = {111},
        pages = {111},
          doi = {10.1088/0004-637X/760/2/111},
archivePrefix = {arXiv},
       eprint = {1209.3772},
 primaryClass = {astro-ph.EP},
       adsurl = {https://ui.adsabs.harvard.edu/abs/2012ApJ...760..111D}
}

@ARTICLE{Dong2019ApJ...870...72D,
       author = {{Dong}, Ruobing and {Liu}, Sheng-Yuan and {Fung}, Jeffrey},
        title = "{Observational Signatures of Planets in Protoplanetary Disks: Planet-induced Line Broadening in Gaps}",
      journal = {\apj},
         year = 2019,
        month = jan,
       volume = {870},
       number = {2},
          eid = {72},
        pages = {72},
          doi = {10.3847/1538-4357/aaf38e},
archivePrefix = {arXiv},
       eprint = {1811.09629},
 primaryClass = {astro-ph.EP},
       adsurl = {https://ui.adsabs.harvard.edu/abs/2019ApJ...870...72D}
}

@software{Dullemond2012ascl.soft02015D,
       author = {{Dullemond}, C.~P. and {Juhasz}, A. and {Pohl}, A. and {Sereshti}, F. and {Shetty}, R. and {Peters}, T. and {Commercon}, B. and {Flock}, M.},
        title = "{RADMC-3D: A multi-purpose radiative transfer tool}",
 howpublished = {Astrophysics Source Code Library, record ascl:1202.015},
         year = 2012,
        month = feb,
          eid = {ascl:1202.015},
archivePrefix = {ascl},
       eprint = {1202.015},
       adsurl = {https://ui.adsabs.harvard.edu/abs/2012ascl.soft02015D}
}

@ARTICLE{Facchini2021AJ....162...99F,
       author = {{Facchini}, Stefano and {Teague}, Richard and {Bae}, Jaehan and {Benisty}, Myriam and {Keppler}, Miriam and {Isella}, Andrea},
        title = "{The Chemical Inventory of the Planet-hosting Disk PDS 70}",
      journal = {\aj},
         year = 2021,
        month = sep,
       volume = {162},
       number = {3},
          eid = {99},
        pages = {99},
          doi = {10.3847/1538-3881/abf0a4},
archivePrefix = {arXiv},
       eprint = {2101.08369},
 primaryClass = {astro-ph.EP},
       adsurl = {https://ui.adsabs.harvard.edu/abs/2021AJ....162...99F}
}

@ARTICLE{Fasano2025A&A...699A.373F,
       author = {{Fasano}, D. and {Benisty}, M. and {Curone}, P. and {Facchini}, S. and {Zagaria}, F. and {Yoshida}, T.~C. and {Doi}, K. and {Sierra}, A. and {Andrews}, S. and {Bae}, J. and {Isella}, A. and {Kurtovic}, N. and {P{\'e}rez}, L.~M. and {Pinilla}, P. and {Rampinelli}, L. and {Teague}, R.},
        title = "{Inner disc and circumplanetary material in the PDS 70 system: Insights from multi-epoch, multi-frequency ALMA observations}",
      journal = {\aap},
         year = 2025,
        month = jul,
       volume = {699},
          eid = {A373},
        pages = {A373},
          doi = {10.1051/0004-6361/202554959},
archivePrefix = {arXiv},
       eprint = {2506.11709},
 primaryClass = {astro-ph.EP},
       adsurl = {https://ui.adsabs.harvard.edu/abs/2025A&A...699A.373F}
}

@ARTICLE{GAIA_2023A&A...674A...1G,
       author = {{Gaia Collaboration} and {Vallenari}, A. and {Brown}, A.~G.~A. and {Prusti}, T. and {de Bruijne}, J.~H.~J. and {Arenou}, F. and {Babusiaux}, C. and {Biermann}, M. and {Creevey}, O.~L. and {Ducourant}, C. and {Evans}, D.~W. and {Eyer}, L. and {Guerra}, R. and {Hutton}, A. and {Jordi}, C. and {Klioner}, S.~A. and {Lammers}, U.~L. and {Lindegren}, L. and {Luri}, X. and {Mignard}, F. and {Panem}, C. and {Pourbaix}, D. and {Randich}, S. and {Sartoretti}, P. and {Soubiran}, C. and {Tanga}, P. and {Walton}, N.~A. and {Bailer-Jones}, C.~A.~L. and {Bastian}, U. and {Drimmel}, R. and {Jansen}, F. and {Katz}, D. and {Lattanzi}, M.~G. and {van Leeuwen}, F. and {Bakker}, J. and {Cacciari}, C. and {Casta{\~n}eda}, J. and {De Angeli}, F. and {Fabricius}, C. and {Fouesneau}, M. and {Fr{\'e}mat}, Y. and {Galluccio}, L. and {Guerrier}, A. and {Heiter}, U. and {Masana}, E. and {Messineo}, R. and {Mowlavi}, N. and {Nicolas}, C. and {Nienartowicz}, K. and {Pailler}, F. and {Panuzzo}, P. and {Riclet}, F. and {Roux}, W. and {Seabroke}, G.~M. and {Sordo}, R. and {Th{\'e}venin}, F. and {Gracia-Abril}, G. and {Portell}, J. and {Teyssier}, D. and {Altmann}, M. and {Andrae}, R. and {Audard}, M. and {Bellas-Velidis}, I. and {Benson}, K. and {Berthier}, J. and {Blomme}, R. and {Burgess}, P.~W. and {Busonero}, D. and {Busso}, G. and {C{\'a}novas}, H. and {Carry}, B. and {Cellino}, A. and {Cheek}, N. and {Clementini}, G. and {Damerdji}, Y. and {Davidson}, M. and {de Teodoro}, P. and {Nu{\~n}ez Campos}, M. and {Delchambre}, L. and {Dell'Oro}, A. and {Esquej}, P. and {Fern{\'a}ndez-Hern{\'a}ndez}, J. and {Fraile}, E. and {Garabato}, D. and {Garc{\'\i}a-Lario}, P. and {Gosset}, E. and {Haigron}, R. and {Halbwachs}, J. -L. and {Hambly}, N.~C. and {Harrison}, D.~L. and {Hern{\'a}ndez}, J. and {Hestroffer}, D. and {Hodgkin}, S.~T. and {Holl}, B. and {Jan{\ss}en}, K. and {Jevardat de Fombelle}, G. and {Jordan}, S. and {Krone-Martins}, A. and {Lanzafame}, A.~C. and {L{\"o}ffler}, W. and {Marchal}, O. and {Marrese}, P.~M. and {Moitinho}, A. and {Muinonen}, K. and {Osborne}, P. and {Pancino}, E. and {Pauwels}, T. and {Recio-Blanco}, A. and {Reyl{\'e}}, C. and {Riello}, M. and {Rimoldini}, L. and {Roegiers}, T. and {Rybizki}, J. and {Sarro}, L.~M. and {Siopis}, C. and {Smith}, M. and {Sozzetti}, A. and {Utrilla}, E. and {van Leeuwen}, M. and {Abbas}, U. and {{\'A}brah{\'a}m}, P. and {Abreu Aramburu}, A. and {Aerts}, C. and {Aguado}, J.~J. and {Ajaj}, M. and {Aldea-Montero}, F. and {Altavilla}, G. and {{\'A}lvarez}, M.~A. and {Alves}, J. and {Anders}, F. and {Anderson}, R.~I. and {Anglada Varela}, E. and {Antoja}, T. and {Baines}, D. and {Baker}, S.~G. and {Balaguer-N{\'u}{\~n}ez}, L. and {Balbinot}, E. and {Balog}, Z. and {Barache}, C. and {Barbato}, D. and {Barros}, M. and {Barstow}, M.~A. and {Bartolom{\'e}}, S. and {Bassilana}, J. -L. and {Bauchet}, N. and {Becciani}, U. and {Bellazzini}, M. and {Berihuete}, A. and {Bernet}, M. and {Bertone}, S. and {Bianchi}, L. and {Binnenfeld}, A. and {Blanco-Cuaresma}, S. and {Blazere}, A. and {Boch}, T. and {Bombrun}, A. and {Bossini}, D. and {Bouquillon}, S. and {Bragaglia}, A. and {Bramante}, L. and {Breedt}, E. and {Bressan}, A. and {Brouillet}, N. and {Brugaletta}, E. and {Bucciarelli}, B. and {Burlacu}, A. and {Butkevich}, A.~G. and {Buzzi}, R. and {Caffau}, E. and {Cancelliere}, R. and {Cantat-Gaudin}, T. and {Carballo}, R. and {Carlucci}, T. and {Carnerero}, M.~I. and {Carrasco}, J.~M. and {Casamiquela}, L. and {Castellani}, M. and {Castro-Ginard}, A. and {Chaoul}, L. and {Charlot}, P. and {Chemin}, L. and {Chiaramida}, V. and {Chiavassa}, A. and {Chornay}, N. and {Comoretto}, G. and {Contursi}, G. and {Cooper}, W.~J. and {Cornez}, T. and {Cowell}, S. and {Crifo}, F. and {Cropper}, M. and {Crosta}, M. and {Crowley}, C. and {Dafonte}, C. and {Dapergolas}, A. and {David}, M. and {David}, P. and {de Laverny}, P. and {De Luise}, F. and {De March}, R.},
        title = "{Gaia Data Release 3. Summary of the content and survey properties}",
      journal = {\aap},
         year = 2023,
        month = jun,
       volume = {674},
          eid = {A1},
        pages = {A1},
          doi = {10.1051/0004-6361/202243940},
archivePrefix = {arXiv},
       eprint = {2208.00211},
 primaryClass = {astro-ph.GA},
       adsurl = {https://ui.adsabs.harvard.edu/abs/2023A&A...674A...1G}
}

@ARTICLE{Garufi2025A&A...694A.290G,
       author = {{Garufi}, A. and {Carrasco-Gonz{\'a}lez}, C. and {Mac{\'\i}as}, E. and {Testi}, L. and {Curone}, P. and {Ricci}, L. and {Facchini}, S. and {Long}, F. and {Manara}, C.~F. and {Pascucci}, I. and {Rosotti}, G. and {Zagaria}, F. and {Clarke}, C. and {Herczeg}, G.~J. and {Isella}, A. and {Rota}, A. and {Mauc{\'o}}, K. and {van der Marel}, N. and {Tazzari}, M.},
        title = "{The centimeter emission from planet-forming disks in Taurus}",
      journal = {\aap},
         year = 2025,
        month = feb,
       volume = {694},
          eid = {A290},
        pages = {A290},
          doi = {10.1051/0004-6361/202452496},
archivePrefix = {arXiv},
       eprint = {2501.11686},
 primaryClass = {astro-ph.EP},
       adsurl = {https://ui.adsabs.harvard.edu/abs/2025A&A...694A.290G}
}

@ARTICLE{Guidi2022A&A...664A.137G,
       author = {{Guidi}, G. and {Isella}, A. and {Testi}, L. and {Chandler}, C.~J. and {Liu}, H.~B. and {Schmid}, H.~M. and {Rosotti}, G. and {Meng}, C. and {Jennings}, J. and {Williams}, J.~P. and {Carpenter}, J.~M. and {de Gregorio-Monsalvo}, I. and {Li}, H. and {Liu}, S.~F. and {Ortolani}, S. and {Quanz}, S.~P. and {Ricci}, L. and {Tazzari}, M.},
        title = "{Distribution of solids in the rings of the HD 163296 disk: a multiwavelength study}",
      journal = {\aap},
         year = 2022,
        month = aug,
       volume = {664},
          eid = {A137},
        pages = {A137},
          doi = {10.1051/0004-6361/202142303},
archivePrefix = {arXiv},
       eprint = {2207.01496},
 primaryClass = {astro-ph.EP},
       adsurl = {https://ui.adsabs.harvard.edu/abs/2022A&A...664A.137G}
}

@ARTICLE{Haffert2019NatAs...3..749H,
       author = {{Haffert}, S.~Y. and {Bohn}, A.~J. and {de Boer}, J. and {Snellen}, I.~A.~G. and {Brinchmann}, J. and {Girard}, J.~H. and {Keller}, C.~U. and {Bacon}, R.},
        title = "{Two accreting protoplanets around the young star PDS 70}",
      journal = {Nature Astronomy},
         year = 2019,
        month = jun,
       volume = {3},
        pages = {749-754},
          doi = {10.1038/s41550-019-0780-5},
archivePrefix = {arXiv},
       eprint = {1906.01486},
 primaryClass = {astro-ph.EP},
       adsurl = {https://ui.adsabs.harvard.edu/abs/2019NatAs...3..749H}
}

@ARTICLE{Hartmann1998ApJ...495..385H,
       author = {{Hartmann}, Lee and {Calvet}, Nuria and {Gullbring}, Erik and {D'Alessio}, Paola},
        title = "{Accretion and the Evolution of T Tauri Disks}",
      journal = {\apj},
         year = 1998,
        month = mar,
       volume = {495},
       number = {1},
        pages = {385-400},
          doi = {10.1086/305277},
       adsurl = {https://ui.adsabs.harvard.edu/abs/1998ApJ...495..385H}
}

@ARTICLE{Hashimoto2012ApJ...758L..19H,
       author = {{Hashimoto}, J. and {Dong}, R. and {Kudo}, T. and {Honda}, M. and {McClure}, M.~K. and {Zhu}, Z. and {Muto}, T. and {Wisniewski}, J. and {Abe}, L. and {Brandner}, W. and {Brandt}, T. and {Carson}, J. and {Egner}, S. and {Feldt}, M. and {Fukagawa}, M. and {Goto}, M. and {Grady}, C.~A. and {Guyon}, O. and {Hayano}, Y. and {Hayashi}, M. and {Hayashi}, S. and {Henning}, T. and {Hodapp}, K. and {Ishii}, M. and {Iye}, M. and {Janson}, M. and {Kandori}, R. and {Knapp}, G. and {Kusakabe}, N. and {Kuzuhara}, M. and {Kwon}, J. and {Matsuo}, T. and {Mayama}, S. and {McElwain}, M.~W. and {Miyama}, S. and {Morino}, J. -I. and {Moro-Martin}, A. and {Nishimura}, T. and {Pyo}, T. -S. and {Serabyn}, G. and {Suenaga}, T. and {Suto}, H. and {Suzuki}, R. and {Takahashi}, Y. and {Takami}, M. and {Takato}, N. and {Terada}, H. and {Thalmann}, C. and {Tomono}, D. and {Turner}, E.~L. and {Watanabe}, M. and {Yamada}, T. and {Takami}, H. and {Usuda}, T. and {Tamura}, M.},
        title = "{Polarimetric Imaging of Large Cavity Structures in the Pre-transitional Protoplanetary Disk around PDS 70: Observations of the Disk}",
      journal = {\apjl},
         year = 2012,
        month = oct,
       volume = {758},
       number = {1},
          eid = {L19},
        pages = {L19},
          doi = {10.1088/2041-8205/758/1/L19},
archivePrefix = {arXiv},
       eprint = {1208.2075},
 primaryClass = {astro-ph.SR},
       adsurl = {https://ui.adsabs.harvard.edu/abs/2012ApJ...758L..19H}
}

@ARTICLE{Hashimoto2015ApJ...799...43H,
       author = {{Hashimoto}, J. and {Tsukagoshi}, T. and {Brown}, J.~M. and {Dong}, R. and {Muto}, T. and {Zhu}, Z. and {Wisniewski}, J. and {Ohashi}, N. and {kudo}, T. and {Kusakabe}, N. and {Abe}, L. and {Akiyama}, E. and {Brandner}, W. and {Brandt}, T. and {Carson}, J. and {Currie}, T. and {Egner}, S. and {Feldt}, M. and {Grady}, C.~A. and {Guyon}, O. and {Hayano}, Y. and {Hayashi}, M. and {Hayashi}, S. and {Henning}, T. and {Hodapp}, K. and {Ishii}, M. and {Iye}, M. and {Janson}, M. and {Kandori}, R. and {Knapp}, G. and {Kuzuhara}, M. and {Kwon}, J. and {Matsuo}, T. and {McElwain}, M.~W. and {Mayama}, S. and {Mede}, K. and {Miyama}, S. and {Morino}, J. -I. and {Moro-Martin}, A. and {Nishimura}, T. and {Pyo}, T. -S. and {Serabyn}, G. and {Suenaga}, T. and {Suto}, H. and {Suzuki}, R. and {Takahashi}, Y. and {Takami}, M. and {Takato}, N. and {Terada}, H. and {Thalmann}, C. and {Tomono}, D. and {Turner}, E.~L. and {Watanabe}, M. and {Yamada}, T. and {Takami}, H. and {Usuda}, T. and {Tamura}, M.},
        title = "{The Structure of Pre-transitional Protoplanetary Disks. II. Azimuthal Asymmetries, Different Radial Distributions of Large and Small Dust Grains in PDS 70}",
      journal = {\apj},
         year = 2015,
        month = jan,
       volume = {799},
       number = {1},
          eid = {43},
        pages = {43},
          doi = {10.1088/0004-637X/799/1/43},
archivePrefix = {arXiv},
       eprint = {1411.2587},
 primaryClass = {astro-ph.SR},
       adsurl = {https://ui.adsabs.harvard.edu/abs/2015ApJ...799...43H}
}

@ARTICLE{Hildebrand2000PASP..112.1215H,
       author = {{Hildebrand}, R.~H. and {Davidson}, J.~A. and {Dotson}, J.~L. and {Dowell}, C.~D. and {Novak}, G. and {Vaillancourt}, J.~E.},
        title = "{A Primer on Far-Infrared Polarimetry}",
      journal = {\pasp},
         year = 2000,
        month = sep,
       volume = {112},
       number = {775},
        pages = {1215-1235},
          doi = {10.1086/316613},
       adsurl = {https://ui.adsabs.harvard.edu/abs/2000PASP..112.1215H}
}

@ARTICLE{Hoang2018ApJ...862..116H,
       author = {{Hoang}, Thiem and {Lan}, Nguyen-Quynh and {Vinh}, Nguyen-Anh and {Kim}, Yun-Jeong},
        title = "{Spinning Dust Emission from Circumstellar Disks and Its Role In Excess Microwave Emission}",
      journal = {\apj},
         year = 2018,
        month = aug,
       volume = {862},
       number = {2},
          eid = {116},
        pages = {116},
          doi = {10.3847/1538-4357/aaccf0},
archivePrefix = {arXiv},
       eprint = {1803.11028},
 primaryClass = {astro-ph.GA},
       adsurl = {https://ui.adsabs.harvard.edu/abs/2018ApJ...862..116H}
}

@ARTICLE{Isella2019ApJ...879L..25I,
       author = {{Isella}, Andrea and {Benisty}, Myriam and {Teague}, Richard and {Bae}, Jaehan and {Keppler}, Miriam and {Facchini}, Stefano and {P{\'e}rez}, Laura},
        title = "{Detection of Continuum Submillimeter Emission Associated with Candidate Protoplanets}",
      journal = {\apjl},
         year = 2019,
        month = jul,
       volume = {879},
       number = {2},
          eid = {L25},
        pages = {L25},
          doi = {10.3847/2041-8213/ab2a12},
archivePrefix = {arXiv},
       eprint = {1906.06308},
 primaryClass = {astro-ph.EP},
       adsurl = {https://ui.adsabs.harvard.edu/abs/2019ApJ...879L..25I}
}

@ARTICLE{Kataoka2014A&A...568A..42K,
       author = {{Kataoka}, Akimasa and {Okuzumi}, Satoshi and {Tanaka}, Hidekazu and {Nomura}, Hideko},
        title = "{Opacity of fluffy dust aggregates}",
      journal = {\aap},
         year = 2014,
        month = aug,
       volume = {568},
          eid = {A42},
        pages = {A42},
          doi = {10.1051/0004-6361/201323199},
archivePrefix = {arXiv},
       eprint = {1312.1459},
 primaryClass = {astro-ph.EP},
       adsurl = {https://ui.adsabs.harvard.edu/abs/2014A&A...568A..42K}
}

@ARTICLE{Kataoka2015ApJ...809...78K,
       author = {{Kataoka}, Akimasa and {Muto}, Takayuki and {Momose}, Munetake and {Tsukagoshi}, Takashi and {Fukagawa}, Misato and {Shibai}, Hiroshi and {Hanawa}, Tomoyuki and {Murakawa}, Koji and {Dullemond}, Cornelis P.},
        title = "{Millimeter-wave Polarization of Protoplanetary Disks due to Dust Scattering}",
      journal = {\apj},
         year = 2015,
        month = aug,
       volume = {809},
       number = {1},
          eid = {78},
        pages = {78},
          doi = {10.1088/0004-637X/809/1/78},
archivePrefix = {arXiv},
       eprint = {1504.04812},
 primaryClass = {astro-ph.EP},
       adsurl = {https://ui.adsabs.harvard.edu/abs/2015ApJ...809...78K}
}

@ARTICLE{Keppler2018A&A...617A..44K,
       author = {{Keppler}, M. and {Benisty}, M. and {M{\"u}ller}, A. and {Henning}, Th. and {van Boekel}, R. and {Cantalloube}, F. and {Ginski}, C. and {van Holstein}, R.~G. and {Maire}, A. -L. and {Pohl}, A. and {Samland}, M. and {Avenhaus}, H. and {Baudino}, J. -L. and {Boccaletti}, A. and {de Boer}, J. and {Bonnefoy}, M. and {Chauvin}, G. and {Desidera}, S. and {Langlois}, M. and {Lazzoni}, C. and {Marleau}, G. -D. and {Mordasini}, C. and {Pawellek}, N. and {Stolker}, T. and {Vigan}, A. and {Zurlo}, A. and {Birnstiel}, T. and {Brandner}, W. and {Feldt}, M. and {Flock}, M. and {Girard}, J. and {Gratton}, R. and {Hagelberg}, J. and {Isella}, A. and {Janson}, M. and {Juhasz}, A. and {Kemmer}, J. and {Kral}, Q. and {Lagrange}, A. -M. and {Launhardt}, R. and {Matter}, A. and {M{\'e}nard}, F. and {Milli}, J. and {Molli{\`e}re}, P. and {Olofsson}, J. and {P{\'e}rez}, L. and {Pinilla}, P. and {Pinte}, C. and {Quanz}, S.~P. and {Schmidt}, T. and {Udry}, S. and {Wahhaj}, Z. and {Williams}, J.~P. and {Buenzli}, E. and {Cudel}, M. and {Dominik}, C. and {Galicher}, R. and {Kasper}, M. and {Lannier}, J. and {Mesa}, D. and {Mouillet}, D. and {Peretti}, S. and {Perrot}, C. and {Salter}, G. and {Sissa}, E. and {Wildi}, F. and {Abe}, L. and {Antichi}, J. and {Augereau}, J. -C. and {Baruffolo}, A. and {Baudoz}, P. and {Bazzon}, A. and {Beuzit}, J. -L. and {Blanchard}, P. and {Brems}, S.~S. and {Buey}, T. and {De Caprio}, V. and {Carbillet}, M. and {Carle}, M. and {Cascone}, E. and {Cheetham}, A. and {Claudi}, R. and {Costille}, A. and {Delboulb{\'e}}, A. and {Dohlen}, K. and {Fantinel}, D. and {Feautrier}, P. and {Fusco}, T. and {Giro}, E. and {Gluck}, L. and {Gry}, C. and {Hubin}, N. and {Hugot}, E. and {Jaquet}, M. and {Le Mignant}, D. and {Llored}, M. and {Madec}, F. and {Magnard}, Y. and {Martinez}, P. and {Maurel}, D. and {Meyer}, M. and {M{\"o}ller-Nilsson}, O. and {Moulin}, T. and {Mugnier}, L. and {Orign{\'e}}, A. and {Pavlov}, A. and {Perret}, D. and {Petit}, C. and {Pragt}, J. and {Puget}, P. and {Rabou}, P. and {Ramos}, J. and {Rigal}, F. and {Rochat}, S. and {Roelfsema}, R. and {Rousset}, G. and {Roux}, A. and {Salasnich}, B. and {Sauvage}, J. -F. and {Sevin}, A. and {Soenke}, C. and {Stadler}, E. and {Suarez}, M. and {Turatto}, M. and {Weber}, L.},
        title = "{Discovery of a planetary-mass companion within the gap of the transition disk around PDS 70}",
      journal = {\aap},
         year = 2018,
        month = sep,
       volume = {617},
          eid = {A44},
        pages = {A44},
          doi = {10.1051/0004-6361/201832957},
archivePrefix = {arXiv},
       eprint = {1806.11568},
 primaryClass = {astro-ph.EP},
       adsurl = {https://ui.adsabs.harvard.edu/abs/2018A&A...617A..44K}
}

@ARTICLE{Keppler2019A&A...625A.118K,
       author = {{Keppler}, M. and {Teague}, R. and {Bae}, J. and {Benisty}, M. and {Henning}, T. and {van Boekel}, R. and {Chapillon}, E. and {Pinilla}, P. and {Williams}, J.~P. and {Bertrang}, G.~H. -M. and {Facchini}, S. and {Flock}, M. and {Ginski}, Ch. and {Juhasz}, A. and {Klahr}, H. and {Liu}, Y. and {M{\"u}ller}, A. and {P{\'e}rez}, L.~M. and {Pohl}, A. and {Rosotti}, G. and {Samland}, M. and {Semenov}, D.},
        title = "{Highly structured disk around the planet host PDS 70 revealed by high-angular resolution observations with ALMA}",
      journal = {\aap},
         year = 2019,
        month = may,
       volume = {625},
          eid = {A118},
        pages = {A118},
          doi = {10.1051/0004-6361/201935034},
archivePrefix = {arXiv},
       eprint = {1902.07639},
 primaryClass = {astro-ph.EP},
       adsurl = {https://ui.adsabs.harvard.edu/abs/2019A&A...625A.118K}
}

@ARTICLE{Kimura2020MNRAS.498.1801K,
       author = {{Kimura}, Hiroshi and {Wada}, Koji and {Kobayashi}, Hiroshi and {Senshu}, Hiroki and {Hirai}, Takayuki and {Yoshida}, Fumi and {Kobayashi}, Masanori and {Hong}, Peng K. and {Arai}, Tomoko and {Ishibashi}, Ko and {Yamada}, Manabu},
        title = "{Is water ice an efficient facilitator for dust coagulation?}",
      journal = {\mnras},
         year = 2020,
        month = oct,
       volume = {498},
       number = {2},
        pages = {1801-1813},
          doi = {10.1093/mnras/staa2467},
archivePrefix = {arXiv},
       eprint = {2008.05841},
 primaryClass = {astro-ph.EP},
       adsurl = {https://ui.adsabs.harvard.edu/abs/2020MNRAS.498.1801K}
}

@ARTICLE{Laor1993ApJ...402..441L,
       author = {{Laor}, Ari and {Draine}, Bruce T.},
        title = "{Spectroscopic Constraints on the Properties of Dust in Active Galactic Nuclei}",
      journal = {\apj},
         year = 1993,
        month = jan,
       volume = {402},
        pages = {441},
          doi = {10.1086/172149},
       adsurl = {https://ui.adsabs.harvard.edu/abs/1993ApJ...402..441L}
}

@ARTICLE{Li2017ApJ...840...72L,
       author = {{Li}, Jennifer I-Hsiu and {Liu}, Hauyu Baobab and {Hasegawa}, Yasuhiro and {Hirano}, Naomi},
        title = "{Systematic Analysis of Spectral Energy Distributions and the Dust Opacity Indices for Class 0 Young Stellar Objects}",
      journal = {\apj},
         year = 2017,
        month = may,
       volume = {840},
       number = {2},
          eid = {72},
        pages = {72},
          doi = {10.3847/1538-4357/aa6f04},
archivePrefix = {arXiv},
       eprint = {1704.06246},
 primaryClass = {astro-ph.SR},
       adsurl = {https://ui.adsabs.harvard.edu/abs/2017ApJ...840...72L}
}

@ARTICLE{Liu2019ApJ...884...97L,
       author = {{Liu}, Hauyu Baobab and {M{\'e}rand}, Antoine and {Green}, Joel D. and {P{\'e}rez}, Sebasti{\'a}n and {Hales}, Antonio S. and {Yang}, Yao-Lun and {Dunham}, Michael M. and {Hasegawa}, Yasuhiro and {Henning}, Thomas and {Galv{\'a}n-Madrid}, Roberto and {K{\'o}sp{\'a}l}, {\'A}gnes and {Takami}, Michihiro and {Vorobyov}, Eduard I. and {Zhu}, Zhaohuan},
        title = "{Diagnosing 0.1-10 au Scale Morphology of the FU Ori Disk Using ALMA and VLTI/GRAVITY}",
      journal = {\apj},
         year = 2019,
        month = oct,
       volume = {884},
       number = {1},
          eid = {97},
        pages = {97},
          doi = {10.3847/1538-4357/ab391c},
archivePrefix = {arXiv},
       eprint = {1908.02981},
 primaryClass = {astro-ph.SR},
       adsurl = {https://ui.adsabs.harvard.edu/abs/2019ApJ...884...97L}
}

@ARTICLE{Li2021SciA....7.3632L,
       author = {{Li}, J. and {Bergin}, E.~A. and {Blake}, G.~A. and {Ciesla}, F.~J. and {Hirschmann}, M.~M.},
        title = "{Earth's carbon deficit caused by early loss through irreversible sublimation}",
      journal = {Science Advances},
         year = 2021,
        month = apr,
       volume = {7},
       number = {14},
        pages = {eabd3632},
          doi = {10.1126/sciadv.abd3632},
archivePrefix = {arXiv},
       eprint = {2104.02702},
 primaryClass = {astro-ph.EP},
       adsurl = {https://ui.adsabs.harvard.edu/abs/2021SciA....7.3632L}
}

@ARTICLE{Liu2019ApJ...877L..22L,
       author = {{Liu}, Hauyu Baobab},
        title = "{The Anomalously Low (Sub)Millimeter Spectral Indices of Some Protoplanetary Disks May Be Explained By Dust Self-scattering}",
      journal = {\apjl},
         year = 2019,
        month = jun,
       volume = {877},
       number = {2},
          eid = {L22},
        pages = {L22},
          doi = {10.3847/2041-8213/ab1f8e},
archivePrefix = {arXiv},
       eprint = {1904.00333},
 primaryClass = {astro-ph.SR},
       adsurl = {https://ui.adsabs.harvard.edu/abs/2019ApJ...877L..22L},
}

@ARTICLE{Liu2022A&A...668A.175L,
       author = {{Liu}, Yao and {Linz}, Hendrik and {Fang}, Min and {Henning}, Thomas and {Wolf}, Sebastian and {Flock}, Mario and {Rosotti}, Giovanni P. and {Wang}, Hongchi and {Li}, Dafa},
        title = "{Underestimation of the dust mass in protoplanetary disks: Effects of disk structure and dust properties}",
      journal = {\aap},
         year = 2022,
        month = dec,
       volume = {668},
          eid = {A175},
        pages = {A175},
          doi = {10.1051/0004-6361/202244505},
archivePrefix = {arXiv},
       eprint = {2210.07478},
 primaryClass = {astro-ph.EP},
       adsurl = {https://ui.adsabs.harvard.edu/abs/2022A&A...668A.175L}
}

@ARTICLE{Liu2024A&A...685A..18L,
       author = {{Liu}, Hauyu Baobab and {Muto}, Takayuki and {Konishi}, Mihoko and {Chung}, Chia-Ying and {Hashimoto}, Jun and {Doi}, Kiyoaki and {Dong}, Ruobing and {Kudo}, Tomoyuki and {Hasegawa}, Yasuhiro and {Terada}, Yuka and {Kataoka}, Akimasa},
        title = "{Forming localized dust concentrations in a dust ring: DM Tau case study. The asymmetric 7 mm dust continuum of the DM Tau disk}",
      journal = {\aap},
         year = 2024,
        month = may,
       volume = {685},
          eid = {A18},
        pages = {A18},
          doi = {10.1051/0004-6361/202348896},
archivePrefix = {arXiv},
       eprint = {2402.02900},
 primaryClass = {astro-ph.EP},
       adsurl = {https://ui.adsabs.harvard.edu/abs/2024A&A...685A..18L}
}

@ARTICLE{Liu2024ApJ...972..163L,
       author = {{Liu}, Hauyu Baobab and {Casassus}, Simon and {Dong}, Ruobing and {Doi}, Kiyoaki and {Hashimoto}, Jun and {Muto}, Takayuki},
        title = "{First JVLA Radio Observation on PDS 70}",
      journal = {\apj},
         year = 2024,
        month = sep,
       volume = {972},
       number = {2},
          eid = {163},
        pages = {163},
          doi = {10.3847/1538-4357/ad5dab},
archivePrefix = {arXiv},
       eprint = {2406.19843},
 primaryClass = {astro-ph.EP},
       adsurl = {https://ui.adsabs.harvard.edu/abs/2024ApJ...972..163L}
}

@ARTICLE{Liu2025SCPMA..6859511L,
       author = {{Liu}, Yao and {Li}, Dafa and {Wang}, Hongchi and {Feng}, Haoran and {Fang}, Min and {Du}, Fujun and {Henning}, Thomas and {Perotti}, Giulia},
        title = "{Dust processing in the terrestrial planet-forming region of the PDS 70 disk}",
      journal = {Science China Physics, Mechanics, and Astronomy},
         year = 2025,
        month = may,
       volume = {68},
       number = {5},
          eid = {259511},
        pages = {259511},
          doi = {10.1007/s11433-024-2597-5},
archivePrefix = {arXiv},
       eprint = {2501.05913},
 primaryClass = {astro-ph.EP},
       adsurl = {https://ui.adsabs.harvard.edu/abs/2025SCPMA..6859511L}
}

@ARTICLE{Long2018ApJ...858..112L,
       author = {{Long}, Zachary C. and {Akiyama}, Eiji and {Sitko}, Michael and {Fernandes}, Rachel B. and {Assani}, Korash and {Grady}, Carol A. and {Cure}, Michel and {Danchi}, William C. and {Dong}, Ruobing and {Fukagawa}, Misato and {Hasegawa}, Yasuhiro and {Hashimoto}, Jun and {Henning}, Thomas and {Inutsuka}, Shu-Ichiro and {Kraus}, Stefan and {Kwon}, Jungmi and {Lisse}, Carey M. and {Liu}, Hauyu Baobab and {Mayama}, Satoshi and {Muto}, Takayuki and {Nakagawa}, Takao and {Takami}, Michihiro and {Tamura}, Motohide and {Currie}, Thayne and {Wisniewski}, John P. and {Yang}, Yi},
        title = "{Differences in the Gas and Dust Distribution in the Transitional Disk of a Sun-like Young Star, PDS 70}",
      journal = {\apj},
         year = 2018,
        month = may,
       volume = {858},
       number = {2},
          eid = {112},
        pages = {112},
          doi = {10.3847/1538-4357/aaba7c},
archivePrefix = {arXiv},
       eprint = {1804.00529},
 primaryClass = {astro-ph.EP},
       adsurl = {https://ui.adsabs.harvard.edu/abs/2018ApJ...858..112L}
}

@ARTICLE{Lyndenbell1974MNRAS.168..603L,
       author = {{Lynden-Bell}, D. and {Pringle}, J.~E.},
        title = "{The evolution of viscous discs and the origin of the nebular variables.}",
      journal = {\mnras},
         year = 1974,
        month = sep,
       volume = {168},
        pages = {603-637},
          doi = {10.1093/mnras/168.3.603},
       adsurl = {https://ui.adsabs.harvard.edu/abs/1974MNRAS.168..603L}
}

@ARTICLE{Muller2018A&A...617L...2M,
       author = {{M{\"u}ller}, A. and {Keppler}, M. and {Henning}, Th. and {Samland}, M. and {Chauvin}, G. and {Beust}, H. and {Maire}, A. -L. and {Molaverdikhani}, K. and {van Boekel}, R. and {Benisty}, M. and {Boccaletti}, A. and {Bonnefoy}, M. and {Cantalloube}, F. and {Charnay}, B. and {Baudino}, J. -L. and {Gennaro}, M. and {Long}, Z.~C. and {Cheetham}, A. and {Desidera}, S. and {Feldt}, M. and {Fusco}, T. and {Girard}, J. and {Gratton}, R. and {Hagelberg}, J. and {Janson}, M. and {Lagrange}, A. -M. and {Langlois}, M. and {Lazzoni}, C. and {Ligi}, R. and {M{\'e}nard}, F. and {Mesa}, D. and {Meyer}, M. and {Molli{\`e}re}, P. and {Mordasini}, C. and {Moulin}, T. and {Pavlov}, A. and {Pawellek}, N. and {Quanz}, S.~P. and {Ramos}, J. and {Rouan}, D. and {Sissa}, E. and {Stadler}, E. and {Vigan}, A. and {Wahhaj}, Z. and {Weber}, L. and {Zurlo}, A.},
        title = "{Orbital and atmospheric characterization of the planet within the gap of the PDS 70 transition disk}",
      journal = {\aap},
         year = 2018,
        month = sep,
       volume = {617},
          eid = {L2},
        pages = {L2},
          doi = {10.1051/0004-6361/201833584},
archivePrefix = {arXiv},
       eprint = {1806.11567},
 primaryClass = {astro-ph.EP},
       adsurl = {https://ui.adsabs.harvard.edu/abs/2018A&A...617L...2M}
}

@ARTICLE{Musiolik2019ApJ...873...58M,
       author = {{Musiolik}, Grzegorz and {Wurm}, Gerhard},
        title = "{Contacts of Water Ice in Protoplanetary Disks{\textemdash}Laboratory Experiments}",
      journal = {\apj},
         year = 2019,
        month = mar,
       volume = {873},
       number = {1},
          eid = {58},
        pages = {58},
          doi = {10.3847/1538-4357/ab0428},
archivePrefix = {arXiv},
       eprint = {1902.08503},
 primaryClass = {astro-ph.EP},
       adsurl = {https://ui.adsabs.harvard.edu/abs/2019ApJ...873...58M}
}

@ARTICLE{Ohashi2020ApJ...900...81O,
       author = {{Ohashi}, Satoshi and {Kataoka}, Akimasa and {van der Marel}, Nienke and {Hull}, Charles L.~H. and {Dent}, William R.~F. and {Pohl}, Adriana and {Pinilla}, Paola and {van Dishoeck}, Ewine F. and {Henning}, Thomas},
        title = "{Solving Grain Size Inconsistency between ALMA Polarization and VLA Continuum in the Ophiuchus IRS 48 Protoplanetary Disk}",
      journal = {\apj},
         year = 2020,
        month = sep,
       volume = {900},
       number = {1},
          eid = {81},
        pages = {81},
          doi = {10.3847/1538-4357/abaab4},
archivePrefix = {arXiv},
       eprint = {2007.15014},
 primaryClass = {astro-ph.EP},
       adsurl = {https://ui.adsabs.harvard.edu/abs/2020ApJ...900...81O}
}

@ARTICLE{Painter2025OJAp....8E.134P,
       author = {{Painter}, Caleb and {Andrews}, Sean M. and {Chandler}, Claire J. and {Ueda}, Takahiro and {Wilner}, David J. and {Long}, Feng and {Macias}, Enrique and {Carrasco-Gonzalez}, Carlos and {Chung}, Chia-Ying and {Liu}, Hauyu Baobab and {Birnstiel}, Tilman and {Hughes}, A. Meredith},
        title = "{Detailed Microwave Continuum Spectra from Bright Protoplanetary Disks in Taurus}",
      journal = {The Open Journal of Astrophysics},
         year = 2025,
        month = sep,
       volume = {8},
          eid = {134},
        pages = {134},
          doi = {10.33232/001c.144268},
archivePrefix = {arXiv},
       eprint = {2507.21268},
 primaryClass = {astro-ph.SR},
       adsurl = {https://ui.adsabs.harvard.edu/abs/2025OJAp....8E.134P}
}

@ARTICLE{Pinilla2024A&A...686A.135P,
       author = {{Pinilla}, Paola and {Benisty}, Myriam and {Waters}, Rens and {Bae}, Jaehan and {Facchini}, Stefano},
        title = "{Survival of the long-lived inner disk of PDS70}",
      journal = {\aap},
         year = 2024,
        month = jun,
       volume = {686},
          eid = {A135},
        pages = {A135},
          doi = {10.1051/0004-6361/202348707},
archivePrefix = {arXiv},
       eprint = {2403.07057},
 primaryClass = {astro-ph.EP},
       adsurl = {https://ui.adsabs.harvard.edu/abs/2024A&A...686A.135P}
}

@ARTICLE{Pollack1994ApJ...421..615P,
       author = {{Pollack}, James B. and {Hollenbach}, David and {Beckwith}, Steven and {Simonelli}, Damon P. and {Roush}, Ted and {Fong}, Wesley},
        title = "{Composition and Radiative Properties of Grains in Molecular Clouds and Accretion Disks}",
      journal = {\apj},
         year = 1994,
        month = feb,
       volume = {421},
        pages = {615},
          doi = {10.1086/173677},
       adsurl = {https://ui.adsabs.harvard.edu/abs/1994ApJ...421..615P}
}

@ARTICLE{Pohl2016A&A...593A..12P,
       author = {{Pohl}, A. and {Kataoka}, A. and {Pinilla}, P. and {Dullemond}, C.~P. and {Henning}, Th. and {Birnstiel}, T.},
        title = "{Investigating dust trapping in transition disks with millimeter-wave polarization}",
      journal = {\aap},
         year = 2016,
        month = aug,
       volume = {593},
          eid = {A12},
        pages = {A12},
          doi = {10.1051/0004-6361/201628637},
archivePrefix = {arXiv},
       eprint = {1607.00387},
 primaryClass = {astro-ph.EP},
       adsurl = {https://ui.adsabs.harvard.edu/abs/2016A&A...593A..12P}
}

@ARTICLE{Rau2011A&A...532A..71R,
       author = {{Rau}, U. and {Cornwell}, T.~J.},
        title = "{A multi-scale multi-frequency deconvolution algorithm for synthesis imaging in radio interferometry}",
      journal = {\aap},
         year = 2011,
        month = aug,
       volume = {532},
          eid = {A71},
        pages = {A71},
          doi = {10.1051/0004-6361/201117104},
archivePrefix = {arXiv},
       eprint = {1106.2745},
 primaryClass = {astro-ph.IM},
       adsurl = {https://ui.adsabs.harvard.edu/abs/2011A&A...532A..71R}
}

@INPROCEEDINGS{Sault1995ASPC...77..433S,
         author = {{Sault}, R.~J. and {Teuben}, P.~J. and {Wright}, M.~C.~H.},
          title = "{A Retrospective View of MIRIAD}",
      booktitle = {Astronomical Data Analysis Software and Systems IV},
           year = 1995,
         editor = {{Shaw}, R.~A. and {Payne}, H.~E. and {Hayes}, J.~J.~E.},
         series = {Astronomical Society of the Pacific Conference Series},
         volume = {77},
          month = jan,
          pages = {433},
  archivePrefix = {arXiv},
         eprint = {astro-ph/0612759},
   primaryClass = {astro-ph},
         adsurl = {https://ui.adsabs.harvard.edu/abs/1995ASPC...77..433S}
  }

@ARTICLE{Sault1996A&AS..117..149S,
       author = {{Sault}, R.~J. and {Hamaker}, J.~P. and {Bregman}, J.~D.},
        title = "{Understanding radio polarimetry. II. Instrumental calibration of an interferometer array.}",
      journal = {\aaps},
         year = 1996,
        month = may,
       volume = {117},
        pages = {149-159},
       adsurl = {https://ui.adsabs.harvard.edu/abs/1996A&AS..117..149S}
}

@ARTICLE{Sierra2025MNRAS.541.3101S,
       author = {{Sierra}, Anibal and {Benisty}, Myriam and {Pinilla}, Paola and {P{\'e}rez}, Laura and {Curone}, Pietro and {Doi}, Kiyoaki and {Facchini}, Stefano and {Fasano}, Daniele and {Andrews}, Sean and {Bae}, Jaehan and {Carpenter}, John and {Czekala}, Ian and {Isella}, Andrea and {Kurtovic}, Nicolas and {Menard}, Francois and {Teague}, Richard},
        title = "{Leaky dust trap in the PDS 70 disc revealed by ALMA Band 9 observations}",
      journal = {\mnras},
         year = 2025,
        month = aug,
       volume = {541},
       number = {4},
        pages = {3101-3112},
          doi = {10.1093/mnras/staf1164},
archivePrefix = {arXiv},
       eprint = {2507.09402},
 primaryClass = {astro-ph.EP},
       adsurl = {https://ui.adsabs.harvard.edu/abs/2025MNRAS.541.3101S}
}

@ARTICLE{Stephens2023Natur.623..705S,
       author = {{Stephens}, Ian W. and {Lin}, Zhe-Yu Daniel and {Fern{\'a}ndez-L{\'o}pez}, Manuel and {Li}, Zhi-Yun and {Looney}, Leslie W. and {Yang}, Haifeng and {Harrison}, Rachel and {Kataoka}, Akimasa and {Carrasco-Gonzalez}, Carlos and {Okuzumi}, Satoshi and {Tazaki}, Ryo},
        title = "{Aligned grains and scattered light found in gaps of planet-forming disk}",
      journal = {\nat},
         year = 2023,
        month = nov,
       volume = {623},
       number = {7988},
        pages = {705-708},
          doi = {10.1038/s41586-023-06648-7},
archivePrefix = {arXiv},
       eprint = {2311.08452},
 primaryClass = {astro-ph.GA},
       adsurl = {https://ui.adsabs.harvard.edu/abs/2023Natur.623..705S}
}

@INPROCEEDINGS{Testi2014prpl.conf..339T,
       author = {{Testi}, L. and {Birnstiel}, T. and {Ricci}, L. and {Andrews}, S. and {Blum}, J. and {Carpenter}, J. and {Dominik}, C. and {Isella}, A. and {Natta}, A. and {Williams}, J.~P. and {Wilner}, D.~J.},
        title = "{Dust Evolution in Protoplanetary Disks}",
    booktitle = {Protostars and Planets VI},
         year = 2014,
       editor = {{Beuther}, Henrik and {Klessen}, Ralf S. and {Dullemond}, Cornelis P. and {Henning}, Thomas},
        month = jan,
        pages = {339-361},
          doi = {10.2458/azu_uapress_9780816531240-ch015},
archivePrefix = {arXiv},
       eprint = {1402.1354},
 primaryClass = {astro-ph.SR},
       adsurl = {https://ui.adsabs.harvard.edu/abs/2014prpl.conf..339T}
}

@ARTICLE{Tazaki2018ApJ...860...79T,
       author = {{Tazaki}, Ryo and {Tanaka}, Hidekazu},
        title = "{Light Scattering by Fractal Dust Aggregates. II. Opacity and Asymmetry Parameter}",
      journal = {\apj},
         year = 2018,
        month = jun,
       volume = {860},
       number = {1},
          eid = {79},
        pages = {79},
          doi = {10.3847/1538-4357/aac32d},
archivePrefix = {arXiv},
       eprint = {1803.03775},
 primaryClass = {astro-ph.EP},
       adsurl = {https://ui.adsabs.harvard.edu/abs/2018ApJ...860...79T}
}

@ARTICLE{Tazaki2019ApJ...885...52T,
       author = {{Tazaki}, Ryo and {Tanaka}, Hidekazu and {Kataoka}, Akimasa and {Okuzumi}, Satoshi and {Muto}, Takayuki},
        title = "{Unveiling Dust Aggregate Structure in Protoplanetary Disks by Millimeter-wave Scattering Polarization}",
      journal = {\apj},
         year = 2019,
        month = nov,
       volume = {885},
       number = {1},
          eid = {52},
        pages = {52},
          doi = {10.3847/1538-4357/ab45f0},
archivePrefix = {arXiv},
       eprint = {1907.00189},
 primaryClass = {astro-ph.EP},
       adsurl = {https://ui.adsabs.harvard.edu/abs/2019ApJ...885...52T}
}

@ARTICLE{Wang2021AJ....161..148W,
       author = {{Wang}, J.~J. and {Vigan}, A. and {Lacour}, S. and {Nowak}, M. and {Stolker}, T. and {De Rosa}, R.~J. and {Ginzburg}, S. and {Gao}, P. and {Abuter}, R. and {Amorim}, A. and {Asensio-Torres}, R. and {Baub{\"o}ck}, M. and {Benisty}, M. and {Berger}, J.~P. and {Beust}, H. and {Beuzit}, J. -L. and {Blunt}, S. and {Boccaletti}, A. and {Bohn}, A. and {Bonnefoy}, M. and {Bonnet}, H. and {Brandner}, W. and {Cantalloube}, F. and {Caselli}, P. and {Charnay}, B. and {Chauvin}, G. and {Choquet}, E. and {Christiaens}, V. and {Cl{\'e}net}, Y. and {Coud{\'e} Du Foresto}, V. and {Cridland}, A. and {de Zeeuw}, P.~T. and {Dembet}, R. and {Dexter}, J. and {Drescher}, A. and {Duvert}, G. and {Eckart}, A. and {Eisenhauer}, F. and {Facchini}, S. and {Gao}, F. and {Garcia}, P. and {Garcia Lopez}, R. and {Gardner}, T. and {Gendron}, E. and {Genzel}, R. and {Gillessen}, S. and {Girard}, J. and {Haubois}, X. and {Hei{\ss}el}, G. and {Henning}, T. and {Hinkley}, S. and {Hippler}, S. and {Horrobin}, M. and {Houll{\'e}}, M. and {Hubert}, Z. and {Jim{\'e}nez-Rosales}, A. and {Jocou}, L. and {Kammerer}, J. and {Keppler}, M. and {Kervella}, P. and {Meyer}, M. and {Kreidberg}, L. and {Lagrange}, A. -M. and {Lapeyr{\`e}re}, V. and {Le Bouquin}, J. -B. and {L{\'e}na}, P. and {Lutz}, D. and {Maire}, A. -L. and {M{\'e}nard}, F. and {M{\'e}rand}, A. and {Molli{\`e}re}, P. and {Monnier}, J.~D. and {Mouillet}, D. and {M{\"u}ller}, A. and {Nasedkin}, E. and {Ott}, T. and {Otten}, G.~P.~P.~L. and {Paladini}, C. and {Paumard}, T. and {Perraut}, K. and {Perrin}, G. and {Pfuhl}, O. and {Pueyo}, L. and {Rameau}, J. and {Rodet}, L. and {Rodr{\'\i}guez-Coira}, G. and {Rousset}, G. and {Scheithauer}, S. and {Shangguan}, J. and {Shimizu}, T. and {Stadler}, J. and {Straub}, O. and {Straubmeier}, C. and {Sturm}, E. and {Tacconi}, L.~J. and {van Dishoeck}, E.~F. and {Vincent}, F. and {von Fellenberg}, S.~D. and {Ward-Duong}, K. and {Widmann}, F. and {Wieprecht}, E. and {Wiezorrek}, E. and {Woillez}, J. and {Gravity Collaboration}},
        title = "{Constraining the Nature of the PDS 70 Protoplanets with VLTI/GRAVITY}",
      journal = {\aj},
         year = 2021,
        month = mar,
       volume = {161},
       number = {3},
          eid = {148},
        pages = {148},
          doi = {10.3847/1538-3881/abdb2d},
archivePrefix = {arXiv},
       eprint = {2101.04187},
 primaryClass = {astro-ph.EP},
       adsurl = {https://ui.adsabs.harvard.edu/abs/2021AJ....161..148W}
}

@ARTICLE{Ueda2024NatAs...8.1148U,
       author = {{Ueda}, Takahiro and {Tazaki}, Ryo and {Okuzumi}, Satoshi and {Flock}, Mario and {Sudarshan}, Prakruti},
        title = "{Support for fragile porous dust in a gravitationally self-regulated disk around IM Lup}",
      journal = {Nature Astronomy},
         year = 2024,
        month = sep,
       volume = {8},
        pages = {1148-1158},
          doi = {10.1038/s41550-024-02308-6},
archivePrefix = {arXiv},
       eprint = {2406.07427},
 primaryClass = {astro-ph.EP},
       adsurl = {https://ui.adsabs.harvard.edu/abs/2024NatAs...8.1148U}
}

@ARTICLE{Vaillancourt2006PASP..118.1340V,
       author = {{Vaillancourt}, John E.},
        title = "{Placing Confidence Limits on Polarization Measurements}",
      journal = {\pasp},
         year = 2006,
        month = sep,
       volume = {118},
       number = {847},
        pages = {1340-1343},
          doi = {10.1086/507472},
archivePrefix = {arXiv},
       eprint = {astro-ph/0603110},
 primaryClass = {astro-ph},
       adsurl = {https://ui.adsabs.harvard.edu/abs/2006PASP..118.1340V}
}

@ARTICLE{VanderMarel2013Sci...340.1199V,
       author = {{van der Marel}, Nienke and {van Dishoeck}, Ewine F. and {Bruderer}, Simon and {Birnstiel}, Til and {Pinilla}, Paola and {Dullemond}, Cornelis P. and {van Kempen}, Tim A. and {Schmalzl}, Markus and {Brown}, Joanna M. and {Herczeg}, Gregory J. and {Mathews}, Geoffrey S. and {Geers}, Vincent},
        title = "{A Major Asymmetric Dust Trap in a Transition Disk}",
      journal = {Science},
         year = 2013,
        month = jun,
       volume = {340},
       number = {6137},
        pages = {1199-1202},
          doi = {10.1126/science.1236770},
archivePrefix = {arXiv},
       eprint = {1306.1768},
 primaryClass = {astro-ph.EP},
       adsurl = {https://ui.adsabs.harvard.edu/abs/2013Sci...340.1199V}
}

@ARTICLE{Weidenschilling1977MNRAS.180...57W,
       author = {{Weidenschilling}, S.~J.},
        title = "{Aerodynamics of solid bodies in the solar nebula.}",
      journal = {\mnras},
         year = 1977,
        month = jul,
       volume = {180},
        pages = {57-70},
          doi = {10.1093/mnras/180.2.57},
       adsurl = {https://ui.adsabs.harvard.edu/abs/1977MNRAS.180...57W}
}

@ARTICLE{Yang2016MNRAS.456.2794Y,
       author = {{Yang}, Haifeng and {Li}, Zhi-Yun and {Looney}, Leslie and {Stephens}, Ian},
        title = "{Inclination-induced polarization of scattered millimetre radiation from protoplanetary discs: the case of HL Tau}",
      journal = {\mnras},
         year = 2016,
        month = mar,
       volume = {456},
       number = {3},
        pages = {2794-2805},
          doi = {10.1093/mnras/stv2633},
archivePrefix = {arXiv},
       eprint = {1507.08353},
 primaryClass = {astro-ph.SR},
       adsurl = {https://ui.adsabs.harvard.edu/abs/2016MNRAS.456.2794Y}
}

@ARTICLE{Yang2020ApJ...889...15Y,
       author = {{Yang}, Haifeng and {Li}, Zhi-Yun},
        title = "{The Effects of Dust Optical Properties on the Scattering-induced Disk Polarization by Millimeter-sized Grains}",
      journal = {\apj},
         year = 2020,
        month = jan,
       volume = {889},
       number = {1},
          eid = {15},
        pages = {15},
          doi = {10.3847/1538-4357/ab5f08},
       adsurl = {https://ui.adsabs.harvard.edu/abs/2020ApJ...889...15Y}
}

@ARTICLE{Yang2025ApJ...989L..43Y,
       author = {{Yang}, Haifeng and {Stephens}, Ian W. and {Lin}, Zhe-Yu Daniel and {Fern{\'a}ndez-L{\'o}pez}, Manuel and {Li}, Zhi-Yun and {Looney}, Leslie W. and {Harrison}, Rachel},
        title = "{Detailed Radial Scale Height Profile of Dust Grains as Probed by Dust Self-scattering in HL Tau}",
      journal = {\apjl},
         year = 2025,
        month = aug,
       volume = {989},
       number = {2},
          eid = {L43},
        pages = {L43},
          doi = {10.3847/2041-8213/adf5b9},
archivePrefix = {arXiv},
       eprint = {2508.01233},
 primaryClass = {astro-ph.EP},
       adsurl = {https://ui.adsabs.harvard.edu/abs/2025ApJ...989L..43Y}
}

@ARTICLE{Yoshida2025ApJ...984L..19Y,
       author = {{Yoshida}, Tomohiro C. and {Curone}, Pietro and {Stadler}, Jochen and {Facchini}, Stefano and {Teague}, Richard and {Momose}, Munetake and {Andrews}, Sean M. and {Bae}, Jaehan and {Barraza-Alfaro}, Marcelo and {Benisty}, Myriam and {Cataldi}, Gianni and {Fasano}, Daniele and {Flock}, Mario and {Fukagawa}, Misato and {Galloway-Sprietsma}, Maria and {Garg}, Himanshi and {Hall}, Cassandra and {Huang}, Jane and {Ilee}, John D. and {Izquierdo}, Andr{\'e}s F. and {Kanagawa}, Kazuhiro and {Lesur}, Geoffroy and {Longarini}, Cristiano and {Loomis}, Ryan A. and {Orihara}, Ryuta and {Pinte}, Christophe and {Price}, Daniel J. and {Rosotti}, Giovanni and {Yen}, Hsi-Wei and {Wafflard-Fernandez}, Gaylor and {Wilner}, David J. and {Winter}, Andrew J. and {W{\"o}lfer}, Lisa and {Zawadzki}, Brianna},
        title = "{exoALMA. XIV. Gas Surface Densities in the RX J1604.3‑2130 A Disk from Pressure-broadened CO Line Wings}",
      journal = {\apjl},
         year = 2025,
        month = may,
       volume = {984},
       number = {1},
          eid = {L19},
        pages = {L19},
          doi = {10.3847/2041-8213/adc42f},
archivePrefix = {arXiv},
       eprint = {2504.19434},
 primaryClass = {astro-ph.EP},
       adsurl = {https://ui.adsabs.harvard.edu/abs/2025ApJ...984L..19Y}
}







   
  



\begin{appendix}




\onecolumn
\nolinenumbers
\section{ALMA Data Reductions}\label{apdx:data}

\subsection{ALMA Data Calibration}
 

For the observations introduced in Section \ref{sec:data}, we downloaded the raw data and followed the standard procedure to perform data calibration using the CASA software package \citep{CASA2022PASP..134k4501C}.
Afterwards, we used the CASA tasks {\tt phaseshift} and {\tt fixplanets} to correct the central position of PDS~70 in all epochs of observations to that on UTC 2023 June 01, assuming that the proper motion \citep{GAIA_2023A&A...674A...1G} is $-29.697$ mas\,yr$^{-1}$ in R.A. and is $-24.041$ mas\,yr$^{-1}$ in Decl., respectively.

For each epoch of observations, we derived the gain-phase self-calibration solutions using aggregated continuum bandwidths and then applied the solutions to individual spectral windows.
The gain-phase self-calibrations for the high angular resolution Band 3 observations  (Table \ref{tab:obs}) and the high angular resolution Band 6 observations at 1287 $\mu$m wavelength failed due to the limited SNR.
Nevertheless, the weather was very good during our observations.
We visually inspected the data during the standard data calibration processes and found that the phase noises are not very large after flagging some bad scans.
Therefore, we do not expect significant losses of intensity due to decoherence. 

\subsection{Imaging of ALMA Observations}

Using the CASA software package, we performed multi-frequency synthesis ({\tt mfs}) imaging \citep{Rau2011A&A...532A..71R} using 1 Taylor coefficient in the spectral model (i.e., {\tt nterms}$=$1).
Due to a realistic consideration of computing power, before performing imaging, we binned the data such that the channel widths are $\lesssim$60 MHz. 
In this case, within the radius of the PDS\,70 disk, frequency smearing leads to much lower than 1\% loss of flux densities at all frequency bands we utilized. 
We created all images using 0$''$.01 pixel sizes. 
We performed primary beam correction for all images. 

\subsubsection{Band 7 polarization images}

We imaged the polarization-calibrated Band~7 data, which were not gain-phase self-calibrated.
In this way, the Stokes {\it I} imaging process was limited by the intensity dynamic ranges owing to the residual phase errors.
As a consequence, the Stokes {\it I} image is noisier than the Stokes {\it Q} and {\it U} images.
This decision stems from the fact that the target source, PDS\,70, exhibits spatially resolved, linearly polarized intensities. 
If we were to use the Stokes {\it I} image for gain-phase self-calibration of the XX and YY observations, it would disrupt the polarization properties of the target source \citep{Sault1996A&AS..117..149S}. 
The effects include artificially suppressing the polarization percentage and rotating the polarization position angles.
In principle, self-calibrating the full polarization data rather than dual polarization data can avoid this effect.
However, the signal-to-noise ratios (SNRs) observed in the XY and YX polarizations were not high enough to support full-polarization gain-phase self-calibration.

We tried Briggs Robust$=$0 and Robust$=$2 weighting and found that only the latter provided high enough SNR for robustly detecting dust polarization.
The Robust$=$2 weighted Stokes {\it I}, {\it Q}, and {\it U} images achieved the synthesized beam of 0$''$.105$\times$0$''$.089 (P.A.=$-$76$^{\circ}$) and the rms noises of 39, 7.6, and 7.5 $\mu$Jy\,beam$^{-1}$, respectively. 

We prepared the polarization intensity, polarization position angle, and polarization percentage images using the CARMA version of MIRIAD software package \citep{Sault1995ASPC...77..433S}.
Polarization percentages and polarization position angles were derived in the regions where the polarization and Stokes {\it I} intensities were detected at $>$3-$\sigma$ and 15-$\sigma$, respectively.
The polarization intensities ({\it PI}) (and the subsequently derived polarization percentages) have been de-biased (i.e,. $PI=\sqrt{Q^2+U^2-\sigma^2}$, where $\sigma$ is the thermal noise; \citealt{Vaillancourt2006PASP..118.1340V}).
The Stokes {\it I}, polarization intensity, polarization percentage, and polarization position angle images prepared this way are shown in Figure \ref{fig:b7pol}.

\subsubsection{Stokes {\it I} images}


We have tried various imaging parameters (e.g., Robust parameters, {\it uv} range, etc). 
In the following, we only introduced those that were directly used in the analyses presented in this paper.

\begin{enumerate}
    \item We imaged the gain-phase self-calibrated Band 7 data using Robust$=$0 weighting and full {\it uv} distance range, to produce the high angular resolution, high fidelity 873 $\mu$m Stokes {\it I} image. This image provides good constraints on the widths of the structures in our MCMC fitting. The achieved synthesized beam and rms noises are 0$''$.064$\times$0$''$.056 (P.A.=$-$49$^{\circ}$) and 12 $\mu$Jy\,beam$^{-1}$, respectively.
    \item Without limiting the {\it uv} distance ranges, we imaged the gain-phase self-calibrated Band 3 (3075 $\mu$m) and Band 4 (2068 $\mu$m) data using Robust$=$2 weighting, and imaged the calibrated Band 6 1287 $\mu$m data and the gain-phase self-calibrated Band 6 1226 $\mu$m data using Robust$=$0 weighting. The achieved synthesized beam sizes and rms noises for these 3075 $\mu$m, 2068 $\mu$m, 1287 $\mu$m, and 1226 $\mu$m images are (0$''$.210$\times$0$''$.145; P.A.=$-$55.6$^{\circ}$; 5.9 $\mu$Jy\,beam$^{-1}$), (0$''$.180$\times$0$''$.164; P.A.=$-$58.8$^{\circ}$; 12 $\mu$Jy\,beam$^{-1}$), (0$''$.104$\times$0$''$.0797; P.A.=$-$84.4$^{\circ}$; 22 $\mu$Jy\,beam$^{-1}$), (0$''$.118$\times$0$''$.104; P.A.=$-$84.2$^{\circ}$; 28 $\mu$Jy\,beam$^{-1}$), respectively. These images, with elliptical synthesized beams, were directly used in our MCMC fittings (see below). For the purpose of diagnosing azimuthally asymmetric substructures, we further smoothed these images to the circular synthesized beams, of which the full width at half maxima (FWHM) are the same as the FWHM of the major axes of the elliptical beams (Figure \ref{fig:maps}).
    \item For the purpose of deriving spectral indices (Figure \ref{fig:spid}), we produced images by limiting the {\it uv} distance ranges to 16--2600 $k\lambda$. We imaged the gain-phase self-calibrated Band 7 data using Robust$=$2 weighting, and imaged the gain-phase self-calibrated Band 6, 1226 $\mu$m data using Robust$=$0 weighting. The former and latter achieved the synthesized beams and rms noise of (0$''$.111$\times$0$''$.097; P.A.=$-$63.7$^{\circ}$; 12 $\mu$Jy\,beam$^{-1}$) and (0$''$.118$\times$0$''$.104; P.A.=$-$84.2$^{\circ}$; 28 $\mu$Jy\,beam$^{-1}$), respectively. We smoothed both images to the common, 0$''$.118$\times$0$''$.104 (P.A.=$-$84.2$^{\circ}$) synthesized beam before measuring spectral indices.
\end{enumerate}

We note that since the high angular resolution Band 6, 1287 $\mu$m data cannot be self-calibrated.
Therefore, the Band 6, 1287 $\mu$m data are not ideal for the purpose of deriving spatially resolved spectral indices. 
In particular, decoherence of signal and smearing of structures due to the residual phase errors can lead to underestimates of intensities at 1287 $\mu$m.
Nevertheless, while the residual phase errors have a more serious impact on the long-baseline data, the measurements of spatially integrated flux densities rely more on the short-spacing observations, which were gain-phase self-calibrated. 
Therefore, the integrated flux densities measured at 1287 $\mu$m are reliable. 

\section{Spectral indices}\label{apdx:spid}

We derived spectral indices ($\alpha$) based on 
\begin{equation}
\log\left(\frac{F_{1}}{F_{2}}\right)=\alpha\log\left(\frac{\nu_{1}}{\nu_{2}}\right),
\label{eq:spid}
\end{equation}
where $F_{1}$ and $F_{2}$ are the flux densities observed at frequency $\nu_{1}$ and $\nu_{2}$, respectively. 
Uncertainties of $\alpha$ were derived considering standard error propagation, while the error ($\sigma$) incorporates the thermal noise $\sigma_{\rm th}$ (i.e., the root-mean-square of the intensities measured in the images) and absolute flux calibration error $\sigma_{\rm abs}$ as $\sigma=\sqrt{ \sigma_{\rm th}^2 + \sigma_{\rm abs}^2 }$.
For each epoch of observations, we assumed that the absolute flux calibration error is a Gaussian random variable. 
We quoted the nominal 1-$\sigma$ absolute flux calibration error (per execution of observation), $\sigma_{\rm abs}^{\rm exc}$, from the ALMA Cycle 11 Technical Handbook\footnote{\url{https://almascience.nrao.edu/proposing/technical-handbook}}.
We adopted $\sigma_{\rm abs}=\sigma_{\rm abs}^{\rm exc}/\sqrt{N^{\rm exc}}$, where $N^{\rm exc}$ is the number of executions of observations that have comparable or better than 0$''$.1 angular resolutions (c.f. Table \ref{tab:obs} and the earlier observational study; \citealt{Facchini2021AJ....162...99F}). 

Our approach slightly overestimated $\sigma$, thereby overestimating the uncertainties of $\alpha$ due to the following two reasons. 
First, the $\sigma_{\rm abs}^{\rm exc}$ is a constant for each epoch of observation, while our formulation treated it as an independent random variable at each independent spatial resolution unit (i.e., Gaussian beam width). 
Second, while $N^{\rm exc}$ only considered the high angular resolution executions of ALMA observations, the short-spacing executions, in fact, can also contribute to noise cancelling. 
The uncertainties of $\alpha_{\rm b6-b7}$ presented in the right panel of Figure \ref{fig:spid} should be regarded as conservative upper limits.
The actual uncertainties of $\alpha_{\rm b6-b7}$ should be smaller, although it is hard to quantify exactly how much smaller it is. 

\section{Radiative Transfer Model}\label{apdx:radmc}

Aiming at addressing the properties of dust that dominates the total intensities and polarization at 873 $\mu$m wavelength, based on MCRT simulations, we constructed a model to compare with the observations on PDS\,70.
This approach is advantageous for our present purpose, since dust scattering is a non-local radiative transfer problem.
Limited by the high computing cost of MCRT simulations, presently, our models for dust density distributions (see below) involve simplifications which help reduce the number of free parameters.
In addition, we avoided including high-density compact (e.g., sub-AU scales) sub-structures which cannot be accurately modeled in the MCRT simulations without using  a prohibitively large number of photons.
Owing to these reasons, we do not attempt to reproduce the detailed intensity distributions at all wavelengths (more in the Caveats and limitations subsection).
Therefore, our fiducial model does not falsify the possibilities of embedding some grown (e.g., $a_{\rm max}>$1000 $\mu$m) dust in the mid-plane of the PDS\,70 disk \citep{Liu2024ApJ...972..163L,Sierra2025MNRAS.541.3101S}. 
The current results show that this approach is promising. 
Future improvements could arise when increased computing power allows for more flexibility in the assumptions regarding dust density distributions.

Similar to the previous studies \citep{Hashimoto2012ApJ...758L..19H,Dong2012ApJ...760..111D,Hashimoto2015ApJ...799...43H}, our model includes a smooth disk that has sub-$\mu$m dust sizes (hereafter small-dust disk; Appendix \ref{subsub:smalldustdisk}), and other dusty structures (i.e., rings and crescents; Appendix \ref{subsub:ring}, \ref{subsub:crescent}) that the $a_{\rm max}$ values are $\gg$1 $\mu$m.
Incorporating the small-dust disk is essential for accurately reproducing the optical and infrared SEDs.
This contributes to building a realistic three-dimensional thermal profile, which is useful for examining the effects of dust self-scattering.
Including the $a_{\rm max}\gg1$ $\mu$m dusty structures is necessary for reproducing the (sub)millimeter intensity distributions resolved in the ALMA observations. 

We note that the previous studies, which were based on lower angular resolution observations, approximated the (sub)millimeter ring using power-law radial density profiles \citep{Hashimoto2012ApJ...758L..19H,Dong2012ApJ...760..111D,Hashimoto2015ApJ...799...43H,Long2018ApJ...858..112L,Keppler2019A&A...625A.118K}.
In our present study, motivated by high angular resolution images (Figure \ref{fig:b7pol}, \ref{fig:maps}; \citealt{Keppler2019A&A...625A.118K,Doi2024ApJ...974L..25D}), we approximated the (sub)millimeter ring using a combination of a truncated narrow Gaussian ring and a truncated wide Gaussian ring.
In addition, we superimposed two dust crescents (crescent-NW, crescent-SW) located in the northwest and southwest of the (sub)millimeter ring, respectively. 
The details of these components on our model will be introduced in the following subsections. 

We based on the following steps to construct the radiative transfer model:

\begin{enumerate}
    \item Using MCRT simulations, we interactively varied the free parameters of the two rings (wide and narrow) and the two crescents (northwest and southwest) until the synthesized (sub)millimeter SED approximate the observed one. Then, we optimize the free parameters of the small-dust disk for reproducing the optical and infrared SED summarized the previous work \citep{Hashimoto2012ApJ...758L..19H}, based on the MCRT simulations and MCMC method. The approach is plausible since the PDS\,70 disk is optically thick at infrared and shorter wavelengths. In this case, small variations of the models for the rings and crescents do not significantly affect the synthesized optical and infrared SED. Conversely, due to the small dust sizes, the small-dust disk makes negligible contribution to the (sub)millimeter intensities. We verified these statements using the MCRT simulations. 
    \item We fixed the free parameters of the small-dust disk, and based on the MCRT simulations and MCMC method to optimize the free parameters for the narrow- and wide-rings for reproducing the intensity distributions resolved over the position angle range 35$^{\circ}$--125$^{\circ}$ in the ALMA 873, 1226, 1287, 2068, and 3075 $\mu$m observations. We selected this position angle range since in our visual inspection the two crescents do not extend into it.
    \item We fixed the free parameters of the small-dust disk and the narrow- and wide-ring, and based on the MCRT simulations and MCMC method to optimize the free parameters for the northwest and southwest crescents. The optimizations were for reproducing the observed intensity distributions over the position angle ranges of 220$^{\circ}$--60$^{\circ}$ and 130$^{\circ}$--215$^{\circ}$, respectively.
\end{enumerate}

When optimizing the parameters for the small-dust disk in Step 1, the log likelihood was derived based on the comparison with the previously published 0.365--90 $\mu$m data \citep{Hashimoto2012ApJ...758L..19H}:

\begin{equation}
    \ln p(F | R_{c}, \beta, h_{\rm 100}) = -\frac{1}{2}\Sigma_{n}\left( \frac{F^{\rm data}_{n} - F^{\rm model}_{n}}{\delta F_{n}}  \right)^{2},
    \label{eq:loglikelihood}
\end{equation}
where $F^{\rm data}_{n}$ and $\delta F_{n}$ are the observed flux density and error of flux density of the n$^{\rm th}$ measurement, and $F^{\rm model}_{n}$ is the flux density of the {\tt RADMC-3D} model at the wavelength of that measurement.

In Steps 2 and 3, the log likelihood was defined as
 \begin{equation}
     \ln p(F | {\rm parms}) = -\frac{1}{2}\Sigma_{n,i}\left( \frac{I^{\rm data}_{n,i} - I^{\rm model}_{n,i}}{\delta I_{n,i}}  \right)^{2},
     \label{eq:loglikelihood_ring}
 \end{equation}
 where parms denote the free parameters mentioned above, $I^{\rm data}_{n,i}$ and $\delta I_{n,i}$ are the observed intensity and error of intensity of the $i^{\rm th}$ pixel of the $n^{\rm th}$ image we compared with, and  $I^{\rm model}_{n,i}$ is the intensity of the {\tt RADMC-3D} model at that pixel and wavelength of the  $n^{\rm th}$ image we compared with.
We evaluated $\ln p(F | {\rm parms})$ from the regions where the intensities (polarization intensity) were detected at $>$5-$\sigma$ ($>$3-$\sigma$), to avoid fitting the low-brightness artifacts that can be produced in the interferometric imaging processes and to avoid the positive biases in the polarization intensities \citep{Vaillancourt2006PASP..118.1340V}. 
We note that although the Stokes {\it Q} and {\it U} images are not positively biased as the polarization intensity image, when the SNR are limited, it is hard to distinguish real negative Stokes {\it Q} and {\it U} intensities from thermal noises and interferometric imaging artifacts. 
Therefore, directly comparing the MCRT model with the observed Stokes {\it Q} and {\it U} intensities still requires truncating the regions that are below 3-$\sigma$ significance. 
In this case, when the polarization position angles are uniform, comparing the Stokes {\it Q} and {\it U} intensities separately and comparing with the polarization intensity are not fundamentally different.
Finally, we masked the inner 0$''$25 projected radius to avoid being confused by the inner disk.

We selected to compare our MCRT model with the following observational images (Appendix \ref{apdx:data})
 \begin{enumerate}
     \item The Robust$=$2 weighted 873 $\mu$m polarization intensity image
     \vspace{-0cm}\item The Robust$=$0  weighted 873 $\mu$m Stokes {\it I} intensity image
     \vspace{-0cm}\item The Robust$=$0 weighted 1226 $\mu$m Stokes {\it I} intensity image
     \vspace{-0cm}\item The Robust$=$0 weighted 1287 $\mu$m Stokes {\it I} intensity image
     \vspace{-0cm}\item The Robust$=$2 weighted 2068 $\mu$m Stokes {\it I} intensity image
     \vspace{-0cm}\item The Robust$=$2 weighted 3075 $\mu$m Stokes {\it I} intensity image
\end{enumerate}
out of the consideration of optimizing the angular resolution versus SNR. 

The computational cost of our approach is high, although it is still manageable. 
The MCMC walkers in all three steps converged in approximately 200 iterations. 
We chose the parameters of our fiducial model to be the 50$^{\rm th}$ percentiles of the MCMC samplers after discarding the burnt-in steps.
The fiducial model successfully reproduced the most prominent features resolved in the ALMA observations.
The results of our fitting are summarized in Table \ref{tab:bestfit}.
Except for $\Sigma_{0}^{\rm narrow-ring}$, the best-fit parameters and the uncertainties were defined by the 16$^{\rm th}$, 50$^{\rm th}$, and 84$^{\rm th}$ percentiles of the MCMC samplers.
We remark that the best-fit $a_{\rm max}^{\rm crescent-NW}$ value is very close to the best-fit value of $a_{\rm max}^{\rm ring}$ in spite of the very different intensities and 873 $\mu$m polarization percentages observed in the ring and in the crescent-NW (Figure \ref{fig:b7pol}).

\begin{figure}
    \centering
    \includegraphics[height=8.5cm]{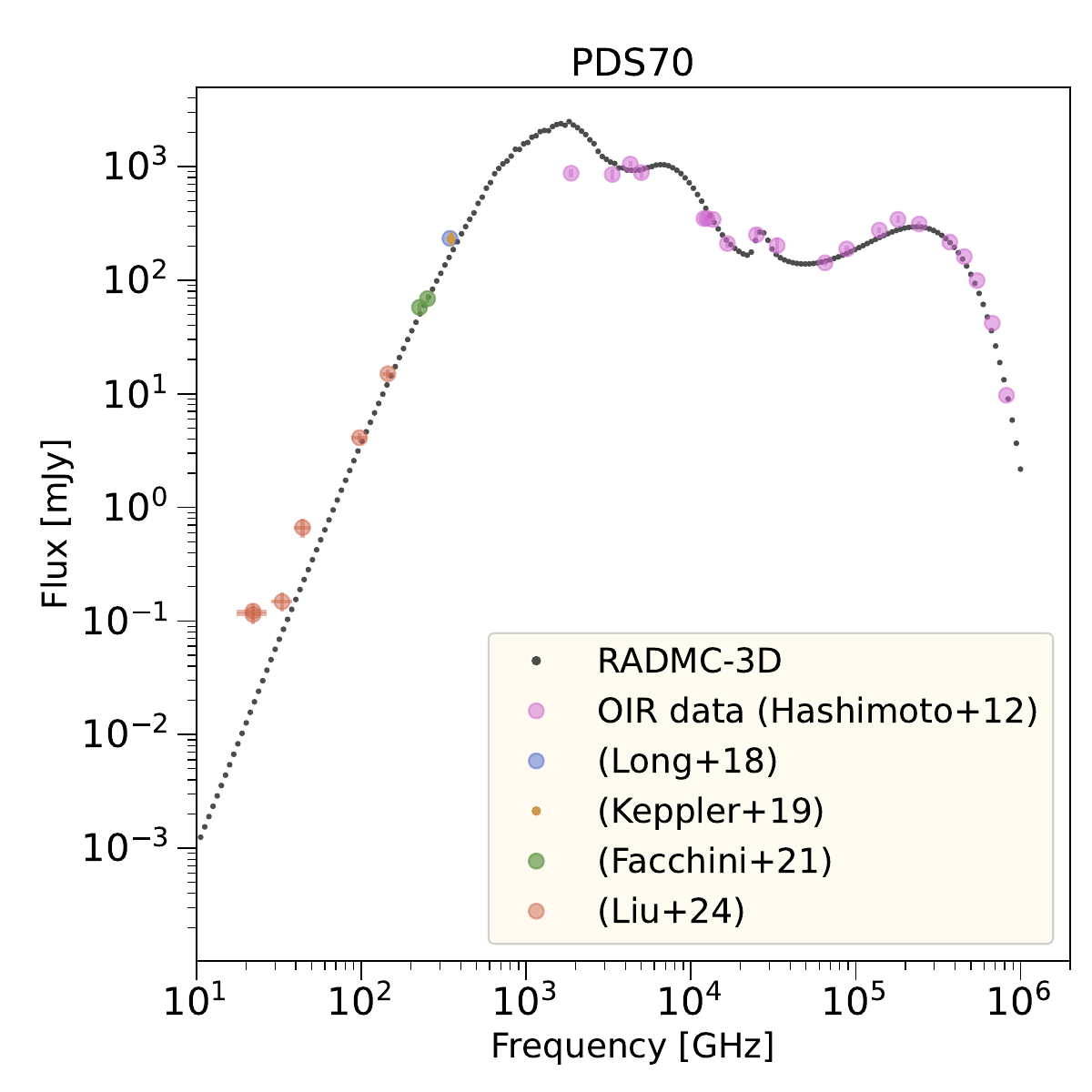}
    \caption{A comparison between the fiducial SED model (black dots; Methods) and observational data (colorful dots; \citealt{Hashimoto2012ApJ...758L..19H,Long2018ApJ...858..112L,Keppler2019A&A...625A.118K,Facchini2021AJ....162...99F,Liu2024ApJ...972..163L}).
    }
    \label{fig:bestfit_sed}
\end{figure}

Figure \ref{fig:bestfit_sed} compares the SED produced from our fiducial model with the observed SEDs.
There is a good consistency between the observed and synthesized SEDs at $<$3000 $\mu$m wavelength (i.e., $>$100 GHz). 
Compared to the observations, the synthesized SED of our model shows excess at $>$3000 $\mu$m wavelength. 
We attribute it to the emission mechanisms that we did not consider in the present model (e.g., free-free emission, or the thermal emission of spinning nanometer-sized dust) and the localized high-density dust concentrations that may have $a_{\rm max}\sim1$ mm, which were discussed in the previous work \citep{Liu2024ApJ...972..163L}.
The free-free emission or the thermal emission of spinning nanometer-sized dust are likely very faint at the wavelength coverages of our ALMA observations \citep{Hoang2018ApJ...862..116H,Chung2025ApJS..277...45C,Garufi2025A&A...694A.290G,Painter2025OJAp....8E.134P}, which should not confuse our modeling \citep{Liu2024ApJ...972..163L}.
The localized high-density dust concentrations are likely situated in the disk mid-plane.
If the localized high-density dust concentrations are in the northwest dust crescent, they should be obscured at 873 $\mu$m wavelengths.
Without considering them will not make an impact on our modelings for the ALMA Band 7 (873 $\mu$m) polarization observations.
However, since the localized high-density dust concentrations can contribute to significant flux densities at long wavelength bands (e.g., 2068--3075 $\mu$m), in terms of SED fittings, ignoring the localized high-density dust concentrations tend to make us slightly overestimate the dust column densities and $a_{\rm max}$ in the rings and crescents.

We remark that, ideally, it is more accurate if we optimize the free parameters of all components (i.e., small-dust, narrow-/wide-rings, crescent-NW/SW) in the model simultaneously.
However, with this approach, considering the realistic computing power, the parameter space is significantly larger than what can be feasibly explored by the MCMC samplers.
Yet the improved accuracy will only yield improvement on some subtle details that are minor to our present science purpose. 

\subsection{Evaluating dust opacity}\label{apdx:opacity}

We evaluated wavelength-dependent dust opacities (including full scattering matrices) using the publicly available {\tt optool} package\footnote{\url{https://github.com/cdominik/optool}}.
We assumed that the size distributions of dust follow $n(a)\propto a^p$ between the minimum and maximum dust grain sizes, $a_{\rm min}$ and $a_{\rm max}$.
Since our main purpose is to model the emission of grown dust that is trapped in the PDS\,70 ring, we assumed $p=-3.5$, which is consistent with fragmentation limited dust growth instead of inward-drift limited dust growth \citep{Birnstiel2016SSRv..205...41B}. 
In general, the $p=-3.5$ assumption provides upper limits of $a_{\rm max}$.
If we make $p$ larger (e.g., $p=-3$ or $p=-2.5$; \citealt{Sierra2025MNRAS.541.3101S}), a larger fraction of overall dust mass will be in the form of grown dust, which in turn leads to a smaller $a_{\rm max}$ value in the fiducial SED model. 

\subsection{Density distribution}\label{apdx:density}

We assumed that the PDS\,70 disk is 51.7$^{\circ}$ inclined and has a 160.1$^{\circ}$ position angle \citep{Keppler2019A&A...625A.118K}. 
We have checked that these parameters can approximate the (sub)millimeter Stokes {\it I} intensity distributions presented in this work (Figure \ref{fig:b7pol}, \ref{fig:maps}). 

\subsubsection{Small-dust disk}\label{subsub:smalldustdisk}
\begin{figure}[h]
    \begin{center}
     \begin{tabular}{cc}
     \includegraphics[width=8cm]{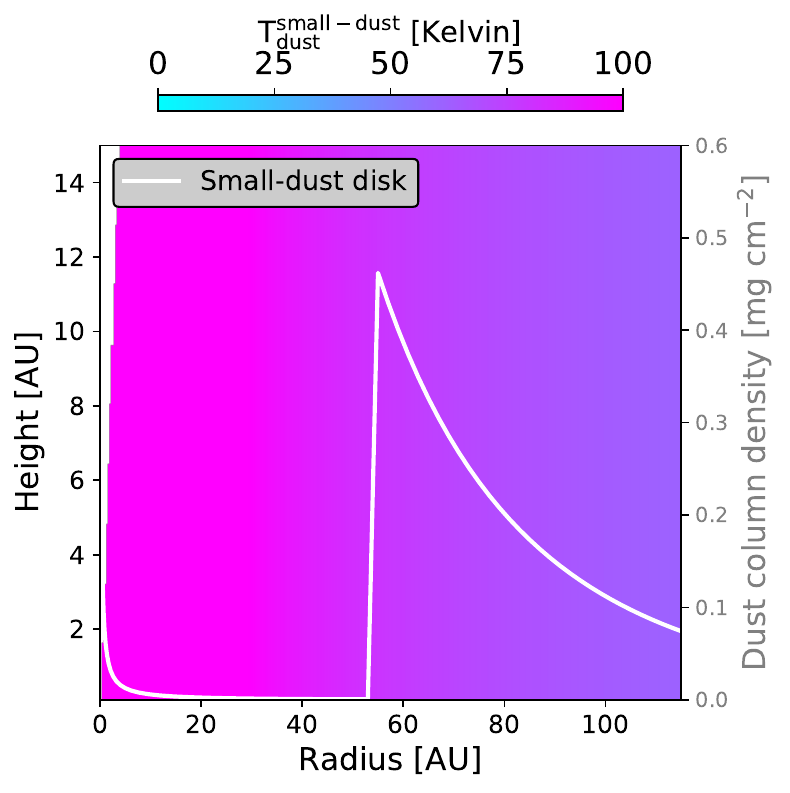} &
     \includegraphics[width=8cm]{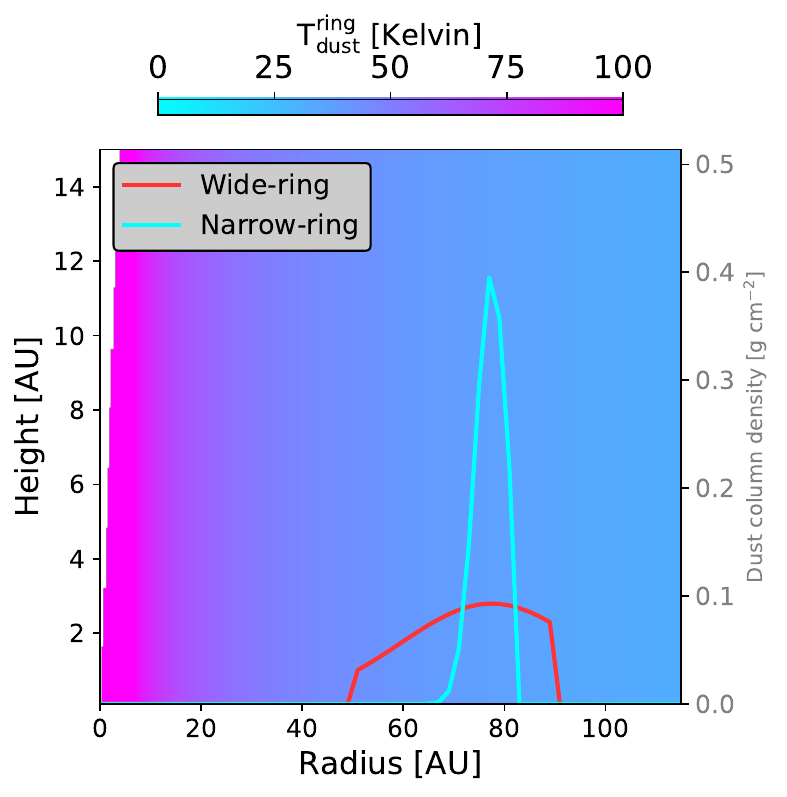} \\
     \end{tabular}
    \end{center}
    \vspace{-0.6cm}
    \caption{Temperature (color) and dust column density (solid lines) profiles of the fiducial model (Methods).
    Left and right panels show these properties of the small-dust disk and the dust rings, respectively.
    Both panels present azimuthally averaged temperatures. 
    }
    \label{fig:profiles}
\end{figure}

Since small dust grains are well coupled with gas, we followed the previous works \citep{Dong2012ApJ...760..111D,Hashimoto2015ApJ...799...43H} to parametrize the (vertically integrated) dust mass surface density ($\Sigma^{\rm small-dust}(R)$) outside the gap radius ($R_{\rm gap}$) in cylindrical-polar coordinates \{R, z\} using a modified formulation for viscous accretion gaseous disk \citep{Lyndenbell1974MNRAS.168..603L,Hartmann1998ApJ...495..385H}:
\begin{equation}
    \Sigma^{\rm small-dust}(R)=\Sigma_{0}^{\rm small-dust}\left(\frac{R}{R_{c}}\right)^{-q}\exp\left[-\left( \frac{R}{R_{c}} \right)^{2-q} \right],
    \label{eq:small-dust}
\end{equation}
where $R_{c}$ is the characteristic radius of the small-dust disk.
Inside $R_{\rm gap}$, we assumed that $\Sigma^{\rm small-dust}(R)$ is $\delta_{\rm cav}$ times smaller than what is described by Equation \ref{eq:small-dust}, where $\delta_{\rm cav}$ is a constant depletion factor.
To reduce the number of free parameters, we assumed $q=1$ following the previous works \citep{Hashimoto2012ApJ...758L..19H,Dong2012ApJ...760..111D,Hashimoto2015ApJ...799...43H}.
We adopted $R_{\rm gap}=$54 AU according to the previous observations \citep{Keppler2018A&A...617A..44K}.
Figure \ref{fig:profiles} shows the resulting $\Sigma^{\rm small-dust}(R)$.

We evaluated the dust mass volume density distribution $\rho(R, z)$ by
\begin{equation}
    \rho(R, z) = \frac{\Sigma^{\rm dust}(R)}{\sqrt{2\pi}h(R)}\exp\left[-\frac{1}{2}\left(\frac{z}{h(R)}\right)^2\right],
    \label{eq:height}
\end{equation}
where $h(R)$ is the characteristic scale height (see below) at radius $R$, and $\Sigma^{\rm dust}(R)=\Sigma^{\rm small-dust}(R)$ for the model of the small-dust disk. 
Following the previous works \citep{Hashimoto2012ApJ...758L..19H,Dong2012ApJ...760..111D,Hashimoto2015ApJ...799...43H}, for the small-dust disk, we adopted $h(R)=h^{\rm small-dust}_{\rm 100}\left(R/100 [AU]\right)^\beta$, where $h^{\rm small-dust}_{\rm 100}$ is the dust scale height at 100 AU radius. We determined the value of $\beta$ using MCMC fittings.

Referencing to the previous studies \citep{Dong2012ApJ...760..111D,Hashimoto2015ApJ...799...43H,Long2018ApJ...858..112L,Keppler2019A&A...625A.118K}, we assumed the total dust mass in the small-dust disk to be 0.330 $M_{\oplus}$; we assumed the dust composition to be 70\% mass in silicate and 30\% mass in graphite \citep{Laor1993ApJ...402..441L}; we assumed $a_{\rm min}^{\rm small-dust}$ and $a_{\rm max}^{\rm small-dust}$ to be 0.001 $\mu$m and 0.15 $\mu$m, respectively.
Since the main purpose for the optimization of the free parameters of the small-dust disk is to establish a realistic thermal profile, the exact dust mass, sizes and composition in the small-dust disk is not important as long as the optical and infrared SED can be reproduced in the synthetic observations. 

Initially, we tried simultaneously determining the values of the four parameters, $R_{c}$, $\beta$, $h^{\rm small-dust}_{\rm 100}$, and $\delta_{\rm cav}$ using MCMC fittings.
We found that there is a degeneracy between the values of $R_{c}$ and $\delta_{\rm cav}$.
Therefore, we re-ran the MCMC fitting with the value of $\delta_{\rm cav}$ fixed to 0.001,  to determine the values of the remaining three free parameters.

\subsubsection{(Sub)millimeter rings}\label{subsub:ring}
Motivated by the $\sim$2000 $\mu$m and $\sim$3000 $\mu$m intensity profiles resolved in the present study and in the previous works \citep{Liu2024ApJ...972..163L,Doi2024ApJ...974L..25D}, we parametrized the dust mass surface density distributions of grown dust grains (Figure \ref{fig:profiles}) in the following ways.
We assumed that the (vertically integrated) dust mass column density profiles of the narrow- and wide-rings can be approximated by the Gaussian distributions for a certain range of radius (see below)
\begin{equation}
    \Sigma^{\rm ring}(R) = \Sigma_{0}^{\rm ring}\exp\left[ - \frac{ (R-R^{\rm ring}_{0})^2 }{ 2(\sigma^{\rm ring})^2 } \right],
    \label{eq:ring}
\end{equation}
were $R_{0}^{\rm ring}$ is the centroid radius of the dusty ring, $\Sigma_{0}^{\rm ring}$ is the dust mass column density at the radius $R=R_{0}^{\rm ring}$ (i.e., the peak dust column density), and $\sigma^{\rm ring}$ is the characteristic width of the ring. 
We defined the inner and outer truncation radii $tr_{\rm in}^{\rm ring}$ and $tr_{\rm out}^{\rm ring}$ and let $\Sigma^{\rm ring}(R> (R_{0}^{\rm ring} + tr_{\rm out}^{\rm ring}) )=0$ and  $\Sigma^{\rm ring}(R< ( R_{0}^{\rm ring} - tr_{\rm in}^{\rm ring} ) )=0$.
Based on MCRT simulations, we found that the truncations are necessary.
Otherwise, the extended rims of the rings will produce too strong far-infrared emission despite the fact that such diffused faint outer rims may not be detectable in the presented ALMA observations. 
Although the sharp truncation appears artificial, limited by angular resolution and sensitivities, its effects are hard to resolve with the present ALMA observations.
In another phrase, the present observations do not well constrain exactly how column density profiles are truncated in the radial direction.

Based on fitting the radial intensity profiles of the ALMA Bands 3, 4, 6, and 7 images, we determined $R_{0}^{\rm narrow-ring}$ to be 77.5 AU. 
For simplicity, we assumed $R_{0}^{\rm wide-ring}=R_{0}^{\rm narrow-ring}$.
In the subsequent MCRT simulations, we verified that this assumption allows producing a model that looks similar to our observations. 

The volume density distributions were evaluated based on an Equation that is analogous to Equation \ref{eq:height}.
We adopted $h^{\rm narrow-ring}(R)=h^{\rm wide-ring}(R)=h^{\rm ring}_{\rm 100}\left(R/100 [AU]\right)^{\beta^{\rm ring} }$.
The value of $\beta^{\rm ring}$ cannot be constrained by the present observations.
Following the previous studies \citep{Hashimoto2012ApJ...758L..19H,Dong2012ApJ...760..111D}, we assumed that the small-dust disk and the embedded grown dust structures have the same indices of scale height (i.e., $\beta^{\rm ring}=\beta$).
The previous study \citep{Dong2012ApJ...760..111D} found that the scale height of the embedded grown dust structures need to be $\sim$10 times smaller than that of the small-dust disk, otherwise, the synthetic far-infrared spectrum will appear over-luminous.  
We verified this by interactively varying $h_{\rm 100}^{\rm narrow-ring}$ and $h_{\rm 100}^{\rm wide-ring}$ in the MCRT simulations.
We found that, as long as $h_{\rm 100}^{\rm ring}$ is smaller than $h_{\rm 100}/10$, the far-infrared spectrum becomes no more sensitive to $h_{\rm 100}^{\rm ring}$. 
The scale heights can also affect the polarization properties at 873 $\mu$m wavelength.
At 873 $\mu$m wavelength, the PDS\,70 (sub)millimeter ring is only marginally optically thick in the vertical direction.
Adopting a large scale height will make the ring less optically thick in the azimuthal direction, which makes the patterns of polarization position angles appear more like the case of a ring rather than that of an inclined disk \citep{Kataoka2015ApJ...809...78K,Yang2016MNRAS.456.2794Y,Yang2025ApJ...989L..43Y}, which is not consistent with the observations (Figure \ref{fig:b7pol}).
On the other hand, adopting a very small scale height will lead to complicated patterns of polarization position angles at the inner and outer rims of the rings \citep{Ohashi2020ApJ...900...81O}, which were not resolved in our observations. 
To avoid the over-luminous synthetic far-infrared spectrum and the (synthetic) pattern of polarization position angles that were not resolved in the ALMA observations, we chose $h_{\rm 100}^{\rm narrow-ring}=h_{\rm 100}^{\rm wide-ring}=$0.45 AU.
The detailed modeling for constraining scale height \citep{Yang2025ApJ...989L..43Y}, which may be confused by the potential azimuthal variation of $a_{\rm max}$ in the specific case of PDS\,70, is beyond the scope of the present paper. 

We assumed the standard DSHARP (water-ice-coated) dust composition \citep{Birnstiel2018ApJ...869L..45B} since both rings are situated outside the $\sim$150 K water snowline (Figure \ref{fig:profiles}). 
We produced the opacity tables using the distribution of hollow spheres (DHS) approach and set the maximum volume fraction $f_{\rm max}$ to 0.8, which is the default of {\tt optool}, to model deviations from perfect spherical symmetry and low-porosity aggregates.
This approach is different from the DSHARP opacity tables released in \citet{Birnstiel2018ApJ...869L..45B} which assumed compact spherical grains.
We checked and found that the differences between our opacity tables and the DSHARP opacity tables are very small.
We assumed the minimum grain sizes $a_{\rm min}^{\rm narrow-ring}=a_{\rm min}^{\rm wide-ring}=0.1$ $\mu$m, while the evaluated dust opacities are not sensitive to the minimum grain sizes.  
Based on numerical simulations, \citet{Pinilla2024A&A...686A.135P} argued that 0.1 $\mu$m is the largest possible $a_{\rm min}$ that can maintain the inner disk of the PDS\,70 system, although there may be model uncertainties.
We tentatively assumed $a_{\rm max}^{\rm narrow-ring}=a_{\rm max}^{\rm wide-ring}=a_{\rm max}^{\rm ring}$, since the present high angular resolution ALMA Bands 3 and 4 observations lack the sensitivity to determine $a_{\rm max}^{\rm narrow-ring}$ and $a_{\rm max}^{\rm wide-ring}$ independently. 
Nevertheless, this assumption is very well motivated by the features of spectral indices and dust polarization described in Section \ref{sec:results}.

The value of $tr_{\rm in}^{\rm wide-ring}$ was chosen such that $R_{0}^{\rm wide-ring}-tr_{\rm in}^{\rm wide-ring}=R_{\rm gap}$, which can make the synthesized infrared images and flux densities consistent with observations.
The intensity distributions at infrared bands are less sensitive to $tr_{\rm out}^{\rm wide-ring}$.
The observations at (sub)millimeter bands do not constrain $tr_{\rm out}^{\rm wide-ring}$ with high precision due to (1) the outer rimg of the wide-ring is faint, and (2) the effect of convolving with the synthesized beam. 
Our present 873 $\mu$m observations significantly detected the PDS~70 ring out to $\sim$100 AU radius.
After some trials with MCRT simulations, we chose $tr_{\rm out}^{\rm wide-ring}=$12 AU, such that $R_{0}^{\rm ring}+tr_{\rm out}^{\rm wide-ring}=$89.5 AU.
In such case, adding one FWHM of a $\sim$0$''$.1 synthesized beam to $R_{0}^{\rm ring}+tr_{\rm out}^{\rm wide-ring}$ will yield a $\sim$100 AU detectable outermost radius. 
We found that this meet the impression of the observations presented in this paper and in the previously published radial intensity profiles \citep{Doi2024ApJ...974L..25D}.
We note that limited by angular resolution, fundamentally, it is hard for our model to discern the errors in the length scales that are smaller than $\sim$0$''$.1/2.35$\sim$4.8 AU (i.e., 1-$\sigma$ beam width).
We based on the MCRT simulations and the MCMC method to optimize the remaining free parameters, $a_{\rm max}^{\rm ring}$, $\Sigma_{0}^{\rm wide-ring}$,  $\sigma^{\rm wide-ring}$, $\Sigma_{0}^{\rm narrow-ring}$,  $\sigma^{\rm narrow-ring}$, $tr_{\rm in}^{\rm narrow-ring}$, and $tr_{\rm out}^{\rm narrow-ring}$.

\subsubsection{Crescents}\label{subsub:crescent}
It is clear that reproducing the northwest crescent in the synthesized (sub)millimeter images requires including an azimuthally asymmetric dusty structure in our model.
In addition, we found that it is not possible to reproduce a (sub)millimeter peak in the south that is offset from the projected major axis, unless we include yet another dusty crescent in the southwest.
The geometry of dusty crescents remain poorly constrained by observations.
Motivated by the previous modeling for the IRS\,48 dusty crescent \citep{VanderMarel2013Sci...340.1199V,Ohashi2020ApJ...900...81O}, we assumed the following two-dimensional dust mass column density profile for certain ranges of radius and deprojected position angle (see below)
\begin{equation}
    \Sigma^{\rm crescent}(R,\phi) = \Sigma_{0}^{\rm crescent}\exp\left[ - \frac{ (R-R^{\rm crescent}_{0})^{p_{r}} }{ 2(\sigma^{\rm crescent})^{p_{r}} } \right]\exp\left[-\frac{(\phi-\phi_{0}^{\rm crescent})^{p_{\phi}} }{2(\sigma^{\rm crescent}_{\phi})^{p_{\phi}}}\right],
    \label{eq:crescent}
\end{equation}
where $\Sigma_{0}^{\rm crescent}$ is the peak dust mass column density, $R_{0}^{\rm crescent}$ is the centroid radius, $\sigma^{\rm crescent}$ is the characteristic width in the radial direction, $\phi_{0}^{\rm crescent}$ is the deprojected centroid position angle, $\phi_{0}^{\rm crescent}$ is the characteristic width in the azimuthal direction (in degree unit), $p_{r}$ is the index of radial density profile, and $p_{\phi}$ is the index of azimuthal density profile. 
We defined the inner and outer truncation radii $tr_{\rm in}^{\rm crescent}$ and $tr_{\rm out}^{\rm crescent}$ and let $\Sigma^{\rm crescent}(R> (R_{0}^{\rm crescent} + tr_{\rm out}^{\rm crescent} ))=0$ and $\Sigma^{\rm crescent}(R< (R_{0}^{\rm crescent} - tr_{\rm in}^{\rm crescent}) )=0$ such that the model does not appear over-luminous at infrared bands.

Motivated by the previous MCRT modeling for the IRS\,48 dusty crescent \citep{VanderMarel2013Sci...340.1199V,Ohashi2020ApJ...900...81O}, we assumed $p_{r}=$4 for the northwest and southwest crescents.
We found that the assumption of a high index of azimuthal density profile (e.g., $p_{\phi}=$4) cannot reproduce the extended northwest crescent resolved in PDS\,70 (Figure \ref{fig:b7pol}, \ref{fig:maps}).
In the present work, We adopted $p_{\phi}=$2.
For both crescents, using MCRT modeling, we found that making $R_{0}^{\rm crescent}$ close to the centroid radius of the ring, 77.5 AU, can yield synthesized (sub)millimeter images that resamble the observed ones. 
Therefore, we fixed the $R_{0}^{\rm crescent}$ of both crescents to 77.5 AU.

We evaluated the volume density distributions based on an equation that is analogous to Equation \ref{eq:height}, assuming that the scale heights do not depend on the azimuth angle $\phi$ and is identical to the scale heights of the narrow- and wide-rings.
The assumptions of dust opacities for the crescents are similar to those for the wide- and narrow-rings (Section \ref{subsub:ring}).

For each crescent, we based on the MCRT simulations and the MCMC method to optimize the remaining free parameters, $a_{\rm max}^{\rm crescent}$, $\Sigma_{0}^{\rm crescent}$, $\sigma^{\rm crescent}$, $tr_{\rm in}^{\rm crescent}$, $tr_{\rm out}^{\rm crescent}$, $\phi_{0}$, and $\sigma_{\phi}^{\rm crescent}$.

\subsection{Radiative  transfer} 
We performed MCRT modeling using the RADMC-3D code (\citealt{Dullemond2012ascl.soft02015D}).
To activate full treatment of dust polarization, we set {\tt scattering\_mode\_max }$=$5. We activated the modified random walk (of photons) by setting {\tt modified\_random\_walk}$=$1.
We note that the default of RADMC-3D is not activating the modified random walk, in case that it lead to less accurate results. 
Nevertheless, we have compared the MCRT simulations with and without activated the modified random walk and found no noticeable differences in the synthetic observations. 
Activating the modified random walk is beneficial for our MCMC fittings since it accelerates the simulations for systems with high optical depths, which is our present case. 
We adopted the RADMC-3D default for the maximum number of dust scattering, {\tt mc\_scat\_maxtauabs}$=$30. 

For each set of chosen free parameters, we first made a thermal run (i.e., using the {\tt mctherm} command) to determine the dust temperatures (Figure \ref{fig:profiles}).
In the thermal runs, the density files simultaneously incorporated the small-dust disk, wide-ring, narrow-ring, northwest crescent, and southwest crescent (Section \ref{subsub:smalldustdisk}, \ref{subsub:ring} \ref{subsub:crescent}).
We found that close to the host protostar where radiation field is strong, the derived temperatures of the small-dust disk can exceed the $\sim$1500 K dust sublimation temperature \citep{Pollack1994ApJ...421..615P}.
To be realistic, after the thermal run, we edited the dust volume density by zeroing it in the regions with $>$1500 K temperature.
Without making this edit, the synthesized optical and infrared SED cannot be consistent with the observed one. 
We produced the synthesized SEDs and images based on the temperature distributions derived from the thermal run and the edited dust volume density.

\subsection{Caveats and limitations}

The purpose of our MCRT model is to verify the plausibility of our current interpretation of the observations of PDS\,70, rather than to precisely reproduce the intensity distributions.
From the point of view of observational data, reproducing the (sub)millimeter intensity distributions resolved in the PDS\,70 disk is subject to a few uncertainties.

Firstly, determining the inclination for an azimuthally asymmetric (sub)millimeter ring is challenging. Earlier studies  \citep{Hashimoto2012ApJ...758L..19H,Dong2012ApJ...760..111D,Hashimoto2015ApJ...799...43H,Long2018ApJ...858..112L} suggested an inclination of 45$^{\circ}$, while later works \citep{Keppler2019A&A...625A.118K,Sierra2025MNRAS.541.3101S} proposed an inclination of 51.7$^{\circ}$. 
Upon visually inspecting and fitting ellipses to the (sub)millimeter images (see Figure \ref{fig:maps}), we found that the inclination likely falls between 45$^{\circ}$ and 51.7$^{\circ}$, at least for certain radius ranges. 
Our MCRT simulations indicate that assuming a constant inclination of 51.7$^{\circ}$ is a reasonable approximation, whereas a constant inclination of 45$^{\circ}$ is inconsistent with the (sub)millimeter interferometric observations.
Although one should bear in mind that a constant inclination in the model is merely an approximation, nevertheless, the range of probable inclinations, either uniform or dependent on radius and azimuthal angle, is small.
We found that changing inclination from 51.7$^{\circ}$ to 45$^{\circ}$ does not yield noticeable changes in the synthesized optical and infrared spectrum. 
Although the small error in the inclination may lead to discrepancies between the synthesized and observed (sub)millimeter intensity distributions, it is unlikely to dramatically change our overall interpretation for the dust sizes and dust mass column densities \citep{Liu2022A&A...668A.175L}.

Second, the observations of the polarization fraction at 873 $\mu$m wavelength and the total intensities at 2068 and 3075 $\mu$m wavelengths are still subject to limited SNR.
This limitation hampers our ability to diagnose the detailed morphologies of the dusty crescents and the spatial dependence of $a_{\rm max}$.
For example, it is still challenging for us to diagnose the $a_{\rm max}$ in the narrow-ring and wide-ring separately. 
Instead, we had to assume that their $a_{\rm max}$ values are comparable.
Moreover, besides the two crescents, there might be more embedded localized dust concentrations which cannot be resolved due to the limited angular resolution and sensitivities of our present ALMA Bands 3 and 4 observations, and due to the high optical depths at shorter wavelength bands \citep{Liu2024A&A...685A..18L}.
This possibility cannot be considered in our radiative transfer modeling, which may lead to discrepancies between the synthesized and observed (sub)millimeter intensity distributions.
This can also confuse the pixel-by-pixel fittings to the (sub)millimeter SEDs.
It is possible to diagnose the embedded dusty concentrations using a multi-components approach in the SED fittings \citep{Li2017ApJ...840...72L,Liu2019ApJ...884...97L,Liu2024A&A...685A..18L,Liu2024ApJ...972..163L}, although it remains difficult to carry out such analyses systematically.
Finally, the infrared spectra of PDS\,70 presented time variabilities \citep{Long2018ApJ...858..112L,Liu2025SCPMA..6859511L}, which led to uncertainties in the derived thermal profiles.

In terms of model, our present approach also involved simplifications.
This is partly due to our compromise to limit computational costs, and partly due to the limited understanding of the system.
For example, we assumed the azimuthally symmetric, power-law profiles for the scale heights of the rings and the crescents, instead of iteratively solving local scale heights based on MCRT simulations. 
The latter approach is not practical due to the prohibitive cost of computing power.
Also, in reality, it is not hard to imagine that the scale height may have dependences on the azimuth angle due to planet-disk interaction.
One should be cautious when applying the scale heights derived in our MCMC fittings to other studies.
In MCRT simulations, errors in the scale heights will lead to errors in the derived thermal profiles, which, in turns, lead to errors in the synthesized intensity distributions. 
Given that the temperature distributions in our MCRT models are relatively smooth (Figure \ref{fig:profiles}), we do not consider this effect significant, in particular, in the Rayleigh-Jeans limit (e.g., frequency bands of the presented ALMA observations).

Our present assumptions of the crescent geometry are not necessarily realistic. 
In addition, it is not necessarily realistic to assume a constant $a_{\rm max}$ value within a crescent \citep{Cazzoletti2018A&A...619A.161C}.
The assumption of position-angle-independent crescent width $\sigma^{\rm crescent}$ (Section \ref{subsub:crescent}) may also be regarded as a simplification. 
The comparison between our model and the observations indicated that a considerable fraction of the dust in the crescents has $a_{\rm max}\lesssim$100 $\mu$m.
It does not rule out the previous claim that some millimeter-sized dust is embedded in the crescents or other unresolved localized substructures \citep{Liu2024ApJ...972..163L}.

Presently, we do not consider the segregation of dust sizes and compositions in the vertical direction.
We think our present approach provides good estimates for the properties of the dust that dominates the continuum luminosity at $\sim$870--3000 $\mu$m wavelengths.
However, there might be a layer of disk atmosphere that the $a_{\rm max}$ value (e.g., $\sim$10 $\mu$m) is in between those of the small-dust disk and the rings and crescents \citep{Sierra2025MNRAS.541.3101S}.
Failing to consider this layer of disk atmosphere may lead to significant errors in the intensity distributions at wavelengths $\lesssim$500 $\mu$m \citep {Sierra2025MNRAS.541.3101S}.
The errors in our synthesized 160 $\mu$m flux density (Section \ref{apdx:mcmc_ring}) may also be partly or largely attributed to this issue. 
On the other hand, constraining the properties of this layer of disk atmosphere requires multi-wavelength observations at $\lesssim$500 $\mu$m wavelengths, which is not available presently.
In this work, we relied on MCRT simulations to derive temperature distributions.
We avoided deriving temperature distributions based on fitting $\sim$400 $\mu$m wavelength observations (e.g., ALMA Band 9 observations), to bypass the systematics caused by the limited understanding of the disk atmosphere. 
Nevertheless, our derived temperature distributions for grown ($\gg$1 $\mu$m) dust appear reasonable as compared to those derived in the SED fittings that included Band 9 observations \citep{Sierra2025MNRAS.541.3101S}.
The disk atmosphere has a small amount of dust mass and should be very optically thin at $>$870 $\mu$m wavelength, which does not significantly affect the scientific discussion of our present work. 

We do not try to model the inner disk in detail.
This is because the inner disk was insufficiently spatially resolved in multi-wavelength observations, and may be confused by non-dust emission mechanisms \citep{Liu2024ApJ...972..163L}.
The models for the inner disk will naturally be very degenerated, therefore lack predicting power. 
The attempts to model it will include many more free parameters, making the MCMC fittings less easy to converge. 
In addition, it makes it more difficult to diagnose the posteriors of the free parameters since the parameter space will have higher dimensions. 

Finally, the uncertainties in dust opacities and dust compositions remain fundamental to all studies that involved fitting (sub)millimeter SEDs, which have been elaborated in some recent studies \citep{Yang2020ApJ...889...15Y,Chung2024ApJS..273...29C,Guidi2022A&A...664A.137G}.
We note that while we adopted the default DSHARP dust opacities presented in \citet{Birnstiel2018ApJ...869L..45B}, based on a population synthesis study, \citet{Delussu2024A&A...688A..81D} suggested that a modified opacity table may be more realistic than the original DSHARP opacities.
However, the population synthesis of \citet{Delussu2024A&A...688A..81D} yielded too low dust spectral indices at 1--3 mm wavelengths to be consistent with the recent surveys of protoplanetary disks (\citealt{Chung2025ApJS..277...45C,Painter2025OJAp....8E.134P}).
Therefore, we do not adopt the updates recommended in \citet{Delussu2024A&A...688A..81D}.
In addition, given the potential segregation of dust sizes and dust compositions, the compositions of small dust constrained by analyzing infrared spectra are not necessarily representative of the compositions of grown dust that resides in the disk mid-plane. 
Moreover, while some observational studies have considered a weakly-porous dust morphology \citep{Stephens2023Natur.623..705S}, we note that evaluating the opacities for porous dust grains remain theoretically challenging and computationally costly \citep{Kataoka2014A&A...568A..42K,Tazaki2018ApJ...860...79T}.
The publicly available codes may not necessarily provide accurate estimates for the opacities of porous dust in the regime of $a_{\rm max}\gtrsim\lambda/2\pi$.
Given that the recent studies favour low porosities \citep{Tazaki2019ApJ...885...52T,Stephens2023Natur.623..705S}, we tentatively assumed a compact dust morphology for simplicity. 
This may lead to underestimates of $a_{\rm max}$ by up to a few times although it is not necessary the case. 

\begin{table}\footnotesize
\caption{Parameters of the fiducial model derived from MCMC fittings}\label{tab:bestfit}

\begin{center}
\begin{tabular}{ccc}
\hline\hline\noalign{\vspace{5pt}}
\multicolumn{3}{c}{Small-dust disk} \\
\noalign{\vspace{2pt}}
\hline\noalign{\vspace{5pt}}
$R_{\rm c}$ & $\beta$ & $h_{\rm 100}$ \\
(AU) & & (AU) \\
\noalign{\vspace{2pt}}
\hline\noalign{\vspace{5pt}}
55.0$^{+1.9}_{-2.0}$ & 1.254$^{+0.008}_{-0.010}$ & 9.14$^{+0.19}_{-0.19}$ \\
\noalign{\vspace{2pt}}\hline
\noalign{\vspace{5pt}}
\end{tabular}
\end{center}

\begin{center}
\begin{tabular}{ccccccc}
\hline\hline\noalign{\vspace{5pt}}
\multicolumn{7}{c}{Narrow-ring and wide-ring} \\
\noalign{\vspace{2pt}}
\hline\noalign{\vspace{5pt}}
$a_{\rm max}^{\rm ring}$         & 
$\Sigma_{0}^{\rm wide-ring}$     &  
$\sigma^{\rm wide-ring}$         & 
$\Sigma_{0}^{\rm narrow-ring}$   &  
$\sigma^{\rm narrow-ring}$       & 
$tr_{\rm in}^{\rm narrow-ring}$  & 
$tr_{\rm out}^{\rm narrow-ring}$ \\
($\mu$m)         &
(g\,cm$^{-2}$)   &
(AU)             &
(g\,cm$^{-2}$)   &
(AU)             &
(AU)             &
(AU)             \\
\noalign{\vspace{2pt}}
\hline\noalign{\vspace{5pt}}
87$^{+10}_{-9}$ &
0.093$^{+0.007}_{-0.015}$ &
18$^{+3}_{-1}$ & 
0.4$^{+0.35}_{-0.01}$ &
3.2$^{+0.1}_{-0.4}$ &
4.9$^{+0.8}_{-0.6}$ &
15$^{+1}_{-1}$ \\
\noalign{\vspace{2pt}}\hline
\noalign{\vspace{5pt}}
\end{tabular}
\end{center}

\begin{center}
\begin{tabular}{ccccccc}
\hline\hline\noalign{\vspace{5pt}}
\multicolumn{7}{c}{Northwest crescent and southwest crescent} \\
\noalign{\vspace{2pt}}
\hline\noalign{\vspace{5pt}}
$a_{\rm max}^{\rm crescent-NW}$   & 
$\Sigma_{0}^{\rm crescent-NW}$    &  
$\sigma^{\rm crescent-NW}$        & 
$tr_{\rm in}^{\rm crescent-NW}$   & 
$tr_{\rm out}^{\rm crescent-NW}$  & 
$\phi_{0}^{\rm crescent-NW}$      &  
$\sigma_{\phi}^{\rm crescent-NW}$ \\
($\mu$m)         &
(g\,cm$^{-2}$)   &
(AU)             &
(AU)             &
(AU)             &
($^{\circ}$)     &
($^{\circ}$)     \\
\noalign{\vspace{2pt}}
\hline\noalign{\vspace{5pt}}
79$^{+44}_{-24}$ & 
11$^{+1}_{-1}$          &
4.0$^{+0.5}_{-0.4}$    &
3.9$^{+0.6}_{-1.0}$    &
6.4$^{+0.2}_{-0.1}$    &
317$^{+1}_{-1}$        &
15.7$^{+0.3}_{-0.2}$  \\
\noalign{\vspace{2pt}}\hline
\noalign{\vspace{5pt}}
$a_{\rm max}^{\rm crescent-SW}$   & 
$\Sigma_{0}^{\rm crescent-SW}$    &  
$\sigma^{\rm crescent-SW}$        & 
$tr_{\rm in}^{\rm crescent-SW}$   & 
$tr_{\rm out}^{\rm crescent-SW}$  & 
$\phi_{0}^{\rm crescent-SW}$      &  
$\sigma_{\phi}^{\rm crescent-SW}$ \\
($\mu$m)         &
(g\,cm$^{-2}$)   &
(AU)             &
(AU)             &
(AU)             &
($^{\circ}$)     &
($^{\circ}$)     \\
\noalign{\vspace{2pt}}
\hline\noalign{\vspace{5pt}}
38$^{+11}_{-11}$ & 
0.15$^{+0.01}_{-0.01}$   &
11$^{+1}_{-1}$    &
7.2$^{+1.0}_{-1.3}$    &
7.8$^{+0.3}_{-0.5}$    &
225$^{+3}_{-3}$        &
12.7$^{+1.5}_{-1.4}$  \\
\noalign{\vspace{2pt}}\hline
\noalign{\vspace{5pt}}
\end{tabular}
\end{center}

%

\end{table}

Given the good fidelity achieved by our ALMA images (Figure \ref{fig:b7pol}, \ref{fig:maps}), we consider the effects mentioned above more important than the effects of the potential interferometric imaging artifacts, which are not trivial to quantitatively assess in an azimuthally asymmetric system. 
In fact, thermal noises in the ALMA Bands 3 and 4 images are likely more important than the interferometric imaging artifacts.
Our present approach of performing modeling in the image instead of the visibility domain benefits from a lower numerical cost, since we do not need to perform Fourier transformation for the MCRT model, and since we can break down the modeling framework into the fittings for a few spatially isolated features.

Instead, the actual dust column density may increase more steeply toward the centroid radius of the narrow-ring than the Gaussian profile. 

\section{Dust fragmentation velocity}\label{apdx:vfrag}

We derived the fragmentation velocity $v_{\rm frag}$ based on the following equation \citep{Birnstiel2016SSRv..205...41B}
\begin{equation}
a_{\rm max}\simeq\frac{\Sigma_{\rm g}}{\rho_{\rm s} \alpha_{\rm vis}}\left(\frac{v_{\rm frag}}{c_{\rm s}}\right)^2,
\end{equation}
where $\alpha_{\rm vis}$ is a dimensionless parameter that characterizes turbulence and viscosity in a viscous accretion disk, $c_{s}=\sqrt{\frac{k_{B}T}{\mu m_{\rm H}}}$ is the thermal sound speed, $\mu=2.33$ is the mean molecular weight, and $m_{\rm H}=1.67\times10^{-24}$ g is the mass of a hydrogen atom. 
We assumed the material density of dust grains $\rho_{\rm s}=$3 g\,cm$^{-2}$ \citep{Birnstiel2016SSRv..205...41B}, and tentatively assumed $\alpha_{\rm vis}=$10$^{-3}$ \citep{Sierra2025MNRAS.541.3101S}; we adopted the temperature $T$ determined from our radiative transfer simulation.
We derived $\Sigma_{g}$ based on the overall dust density in our model, assuming a gas-to-dust mass ($g$) ratio of 100 \citep{Yoshida2025ApJ...984L..19Y}.

Since we do not consider dust porosity, the assumed value of $\rho_{\rm s}$ may be overestimated by up to 1 order of magnitude \citep{Tazaki2019ApJ...885...52T,Stephens2023Natur.623..705S}, while we might underestimated $a_{\rm max}$ by a few times.
These effects of overestimating $\rho_{s}$ and underestimating $a_{\rm max}$ compensate for each other in the derivation of $v_{\rm frag}$.
Therefore, we do not think the assumption of a compact instead of slightly (e.g., $\sim$90\%) porous grain morphology will significantly bias the derivation of $v_{\rm frag}$.

Assuming $g=$100 may overestimated $\Sigma_{\rm g}$, since the gas-to-dust mass ratio may be lowered in dust traps \citep{Birnstiel2016SSRv..205...41B}.
This leads to underestimates of $v_{\rm frag}$. 
This may be more concerning for the dust narrow-ring and the dust crescents, as these structures could represent dust traps that have lower $g$ values (Figure \ref{fig:best-fit}).
Conversely, if a significant fraction of dust mass is removed from the wide-ring due to dust migration, the assumption of $g=$100 may underestimated $\Sigma_{\rm g}$ in the wide-ring, leading to overestimates of $v_{\rm frag}$.
For this reason, we avoided discussing $v_{\rm frag}$ inside the (sub)millimeter cavity as it would be significantly overestimated.

Planet-disk interaction may induce turbulence \citep{Bi2021ApJ...912..107B,Dong2019ApJ...870...72D}, which effectively enhances the value of $\alpha_{\rm vis}$.
Without considering this mechanism may lead to underestimates of $\alpha_{\rm vis}$, thereby underestimating $v_{\rm frag}$.
Nevertheless, our presently assumed $\alpha_{\rm vis}$ value is already relatively high.
It is not very likely that our assumed $\alpha_{\rm vis}$ value is significantly underestimated.

\end{appendix}
\end{document}